\documentclass[screen,acmtog,nonacm]{acmart}

\usepackage{amsmath}
\usepackage{bm}
\usepackage{color,balance,microtype,booktabs}
\usepackage[fleqn,tbtags]{mathtools}
\usepackage{hyperref}
\usepackage[capitalise]{cleveref} % should be placed after the hyperref package
\usepackage[detect-weight]{siunitx}
\usepackage{multirow}
\usepackage{multicol}
\usepackage{enumitem}
\usepackage{stfloats}
\usepackage{textcomp}
\usepackage{eso-pic}

\usepackage[english]{babel}
\usepackage{amsthm}

\usepackage{amsmath,amssymb}

\graphicspath{{./images/}}

\def \etal {{\emph{et al}.\thinspace}}

\newcommand\dif{\mathop{}\!\mathrm{d}}

\usepackage{xcolor}
\usepackage{pifont}

\AtBeginDocument{%
    }

\makeatletter
\@namedef{ver@everyshi.sty}{}
\makeatother

\begin{document}
\title{Lattice Structure Optimization for Additive Manufacturing: Manufacturability-Driven Design and Pareto Front Construction}
% \titlenote{Project page:\url{https://xing-yuu.github.io/project/MAPLE/MAPLE.html}}

\author{Yu Xing}
\affiliation{%
    \institution{Shandong University}
    \city{Qingdao}
    \country{China}
}
\email{xing\_yu@mail.sdu.edu.cn}

\author{Yang Liu}
\affiliation{%
    \institution{Microsoft Research Asia}
    \city{Beijing}
    \country{China}
}
\email{yangliu@microsoft.com}

\author{Lin Lu}
\authornote{Corresponding author}
\affiliation{%
    \institution{Shandong University}
    \city{Qingdao}
    \country{China}
}
\email{llu@sdu.edu.cn}
\titlenote{Project page: \url{https://xing-yuu.github.io/project/MAPLE/MAPLE.html}}
\begin{abstract}
Lattice metamaterials provide an important foundation for lightweight and multifunctional structural design, while additive manufacturing enables the fabrication of complex lattice geometries. However, multiphysics lattice unit-cell design for additive manufacturing still faces two challenges. First, existing methods struggle to efficiently construct multi-objective Pareto fronts with adequate coverage under limited computational budgets. Second, manufacturing constraints, including overhangs, enclosed cavities, and restricted powder-removal channels, reduce the feasible design space. To address these challenges, a method driven by manufacturing constraints is proposed for lattice structure optimization and Pareto front construction. First, manufacturing constraints are incorporated into topology optimization in a differentiable form within an inverse homogenization framework, allowing physical performance and manufacturability to be considered within the same optimization process. Then, a progressive Pareto front construction mechanism is developed. 
A density generation network learns latent representations of high-quality lattice structures; the latent representations of neighboring nondominated solutions are interpolated, and the decoded density fields are used to initialize subsequent topology optimization. Newly obtained nondominated solutions are used to update the network and the sample set, progressively improving the coverage of the sampled manufacturable nondominated set. Experiments on three-dimensional periodic lattice unit cells show that, under an identical budget of 1000 optimization runs, the network initialization strategy achieves an optimization success rate of 92.60\%, compared with 78.30\% for random initialization. It yields 916 manufacturable samples, compared with 776 obtained through random initialization. The Pareto front obtained through network initialization has a hypervolume of 0.0787, exceeding the value of 0.0675 obtained through random initialization. These results demonstrate that the proposed method can efficiently construct Pareto fronts for multiple physical properties while accounting for manufacturing constraints and obtain manufacturable nondominated solution sets with broader coverage. 
\textbf{This manuscript is an English translation of a Chinese article accepted for publication in the Journal of Computer-Aided Design \& Computer Graphics (2026): }\href{https://www.jcad.cn/article/doi/10.3724/SP.J.1089.2026-00157}{DOI: 10.3724/SP.J.1089.2026-00157}.
\end{abstract}

\maketitle

\authorsaddresses{Xu Xing, xing_
yu@mail.sdu.edu.cn; Yang Liu, yangliu@
microsoft.com; Lin Lu, llu@sdu.edu.cn}

\section{Introduction}

Lattice metamaterials use microscale geometry to tailor macroscopic properties and thus provide an important route to lightweight and multifunctional structural design~\cite{li1996microstructure,fleck2010micro}. This geometry-driven mechanism enables high specific stiffness, high specific strength, and tunable anisotropy~\cite{schumacher2015microstructures}. Advances in additive manufacturing (AM) now allow complex periodic microstructures to be fabricated with high precision~\cite{askari2020additive,kulagin2020architectured}, shifting the central question from whether such structures can be manufactured to how their performance can be optimized under manufacturing constraints~\cite{ccalicskan2026design,akula2026metallic}.

Inverse homogenization establishes a mapping between microscale topology and macroscale effective properties and is therefore a core tool for this problem~\cite{dai2025,andreassen2014determine,liu2025ms,ji2026designing,sigmund1994materials,duan2025inverse}. Considerable progress has been made in the optimization of elastic, thermally conductive, and multifunctional microstructures~\cite{osanov2016topology,zhang2023optimized,chen2026designing,xue2025mind,duan2025inverse,yang2026guided,zheng2023unifying}. A major limitation, however, is that most existing methods focus on a single property or reduce multiple objectives to a weighted scalar objective~\cite{krysko2019topological,meng2025topology}. Consequently, they cannot fully reveal the global trade-offs among mechanical performance, thermal conductivity, lightweight design, and manufacturability~\cite{shi2023process,yang2024multifunctional}. Multiphysics lattice design generally has no unique optimum. Designers instead require a diverse set of mutually nondominated and manufacturable candidates whose objective-space distribution forms a Pareto front, from which a design can be selected for a specific application.

Achieving this goal raises two major challenges.

First, how can manufacturing constraints be incorporated directly into optimization? Although AM greatly expands the range of realizable lattice geometries, manufacturing limitations remain consequential. In metal powder-bed fusion, for example, local overhangs, enclosed cavities, and excessively narrow powder-removal channels can lead to build failure or trapped powder~\cite{langelaar2016topology,zhang2022structural,xiong2020new,liu2022topology}. A posteriori manufacturability checks or geometric corrections often alter the original topology, volume fraction, and effective physical properties, and they do not capture how manufacturing constraints restrict the feasible region. Existing treatments, including minimum feature-size constraints, overhang-angle constraints, and self-support filters~\cite{gaynor2016topology,guest2004achieving,zhou2015minimum,langelaar2016topology,qian2017undercut,allaire2017structural}, primarily control local geometry and lack a unified representation of global connectivity and powder-removal paths in periodic lattices. Moreover, local manufacturability of an isolated unit cell does not guarantee manufacturability after periodic tiling. Manufacturing constraints determine not only the feasibility of an individual structure but also the attainable trade-off boundary among multiple objectives.

Second, how can a sufficiently representative Pareto front be constructed within a limited computational budget? Inverse-homogenization problems are high-dimensional, strongly nonconvex, and sensitive to the initial field~\cite{sigmund1998numerical,alvarez2019influence}. Conventional weight sweeps or random initializations are computationally expensive and often leave the front incompletely covered~\cite{marler2010weighted,kim2005adaptive}. Rather than relying solely on independent samples, the front must be actively guided and progressively improved within the constrained feasible space.

We address these challenges with a manufacturability-constrained lattice optimization framework for AM and a progressive strategy for Pareto-front construction. Within an inverse-homogenization formulation, effective mechanical and thermal properties are modeled jointly, while automatic-differentiation-compatible, piecewise-differentiable penalties for overhangs, enclosed cavities, and minimum powder-removal channels are embedded directly in topology optimization. Performance and manufacturability are therefore handled within the same optimization process. Pareto-front construction is formulated as an iterative process. A target-performance-conditioned density generation network encodes the design condition and target properties into a latent representation. Neighboring nondominated solutions in sparsely covered front regions are interpolated in latent space and decoded into initial density fields. Because the latent space is learned from high-quality solutions, the decoded fields provide priors near the observed trade-off boundary. Newly obtained structures are added to the dataset and used to update the network, forming a closed loop between front evolution and model updating that progressively improves the coverage of manufacturable nondominated solutions under a fixed budget.

The main contributions are as follows:

\begin{enumerate}[leftmargin=*]\setlength\itemsep{0mm}
    \item We formulate multiphysics lattice design for AM as the construction of a manufacturable nondominated set. Effective mechanical properties, thermal conductivity, and volume fraction are evaluated consistently within inverse homogenization, allowing their attainable trade-offs to be analyzed under the same manufacturing assumptions.
    \item We formulate automatic-differentiation-compatible manufacturing penalties for periodic lattices. Continuous-density terms represent local overhangs, enclosed cavities, and restricted powder-removal channels in metal powder-bed fusion, allowing these nonmanufacturable features to be suppressed during optimization.
    \item We develop a network-guided progressive Pareto-front construction strategy. A target-performance-conditioned density generation network learns priors from high-quality lattices, and latent interpolation between neighboring nondominated solutions generates new initial fields. Manufacturability-constrained topology optimization, nondominated filtering, and network updating then form a closed loop that yields a more complete set of manufacturable Pareto solutions within a limited computational budget.
\end{enumerate}

\section{Related Work}

\subsection{Lattice Design and Inverse Homogenization}

Numerical homogenization is a fundamental tool for connecting unit-cell geometry to macroscopic effective properties. By imposing periodic boundary conditions on a representative unit cell, effective elastic and thermal-conductivity tensors can be computed, thereby mapping a microscale topology to quantifiable macroscale parameters~\cite{andreassen2014determine,dong2019149}. Inverse homogenization uses this mapping in the opposite direction, searching for a unit-cell topology that satisfies the prescribed target properties. It has become a classical strategy for designing lattice metamaterials~\cite{sigmund1994materials}.

Extensive studies have developed this framework for elasticity, thermal conduction, anisotropy, and related properties~\cite{osanov2016topology,ji2026designing}. Zhang \etal developed a GPU-parallel solver for large-scale three-dimensional problems~\shortcite{zhang2023optimized}; 
Xing \etal introduced a geometric-multigrid-aligned transformer with numerical refinementfor high-fidelity mechanical and thermal microstructure homogenization~\shortcite{xing2026gmt}. Chen \etal optimized high-thermal-conductivity microstructures~\shortcite{chen2026designing}; and Xue \etal proposed an inverse-design method based on neural representations~\shortcite{xue2025mind}. Generative models have also been introduced to learn property-to-structure mappings and accelerate the generation of candidate lattices~\cite{duan2025inverse,yang2026guided,zheng2023unifying}.

Nevertheless, most of these methods are restricted to a single property or a fixed design condition. Effective tools for jointly designing multiphysics properties, particularly for systematically characterizing their trade-offs, remain limited. This motivates a multiobjective formulation within inverse homogenization in which the competition and compromise among properties are represented by a set of nondominated solutions.

\subsection{Manufacturability-Constrained Lattice Optimization for Additive Manufacturing}

Additive manufacturing makes complex lattice geometries realizable, but manufacturability remains a major barrier to engineering deployment~\cite{thompson2016design,liu2018current,ccalicskan2026design}. In metal powder-bed fusion and related processes, geometry and fabrication are tightly coupled. Slender members may fall below the process resolution; local overhangs can require supports and cause build defects~\cite{gaynor2016topology}; and enclosed cavities or narrow pore channels can trap unfused powder~\cite{wang2022topology}, complicating post-processing and reducing in-service reliability.

Several approaches have been proposed to improve printability. Guest \etal~\shortcite{guest2004achieving} and Zhou \etal~\shortcite{zhou2015minimum} controlled minimum feature size through geometric constraints. Gaynor and Guest~\shortcite{gaynor2016topology} and Langelaar~\shortcite{langelaar2016topology} incorporated overhang and self-support constraints into optimization. Qian~\shortcite{qian2017undercut} and Allaire \etal~\shortcite{allaire2017structural} addressed overhangs from the perspectives of density gradients and shape optimization, respectively. Xiong \etal~\shortcite{xiong2020new} and Liu \etal~\shortcite{liu2022topology} explored the elimination of enclosed cavities and the design of powder-removal channels.

These methods primarily control local geometric features and do not jointly describe global connectivity and powder-removal paths for periodic lattices. This gap motivates incorporating manufacturing constraints directly into topology optimization so that the feasible design region and its effect on multiobjective trade-offs can be assessed together.

\subsection{Multiobjective Optimization and Pareto-Front Construction for Multiphysics Lattices}

A Pareto front describes multiphysics trade-offs more completely than weighted scalarization because it represents an entire limiting boundary rather than a single compromise~\cite{de2023multi,garland2021pragmatic}. Constructing such a front is difficult, however, because inverse-homogenization topology optimization is high-dimensional, nonconvex, and sensitive to initial conditions~\cite{sigmund1998numerical,alvarez2019influence}. Repeated weight sampling and random initialization remain common, but they incur high computational costs and often cover the front unevenly~\cite{marler2010weighted,kim2005adaptive}. Manufacturing constraints further contract the feasible design space, making candidates that satisfy both performance and manufacturability requirements even harder to obtain. These difficulties motivate guided strategies that improve the search efficiency and distribution quality of nondominated solutions under a limited computational budget.

Deep learning has recently been used for multiobjective inverse design and Pareto-front construction. Hu \etal combined a machine-learning surrogate with NSGA-II to construct an acoustic-absorption--energy-absorption Pareto set for shell lattices~\shortcite{Hu31122024}. Frey \etal used graph neural networks and latent-space optimization for multiobjective inverse design of multimaterial truss lattices~\shortcite{frey2025multi}. These studies show that data-driven surrogates and learned structural priors can reduce the cost of searching a multiobjective design space. In contrast, our method uses the output of a target-performance-conditioned network as the initial field for manufacturability-constrained inverse-homogenization topology optimization. Newly obtained manufacturable nondominated solutions are continually returned to the training set, progressively expanding the Pareto front through a closed loop between front evolution and model updating.

Building on manufacturability-constrained optimization, the proposed data-driven iterative strategy interpolates nondominated solutions in the learned latent space to generate new initializations for optimization. It therefore couples structural-prior learning with Pareto-front evolution to improve both the efficiency and quality of multiphysics front construction.
\section{Method}
\begin{figure}[t]
\centering
\includegraphics[width=\linewidth]{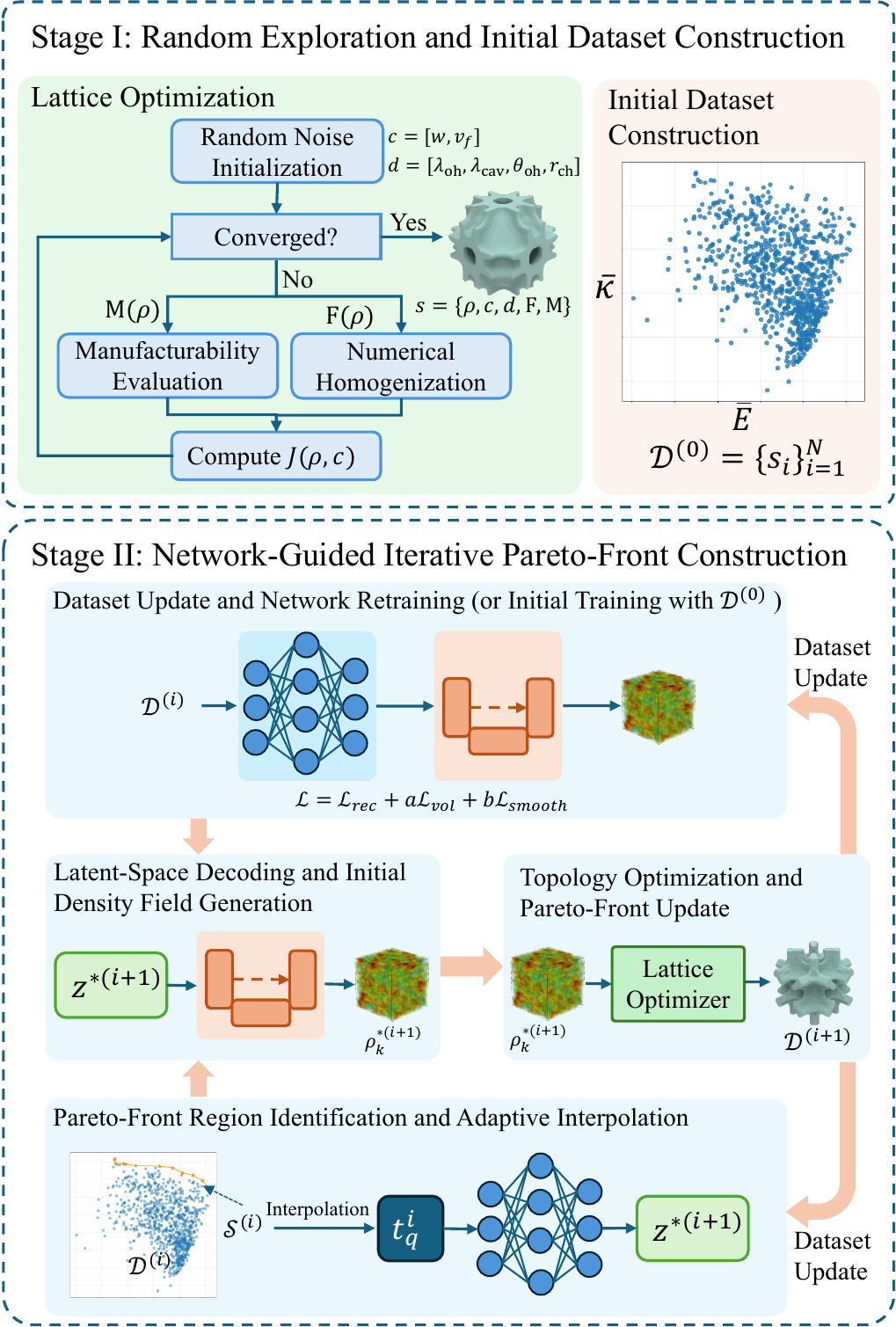}
\caption{\textbf{Framework for manufacturability-constrained lattice optimization and progressive Pareto-front construction.} Stage I generates manufacturable candidates by inverse-homogenization topology optimization under manufacturing constraints. Stage II progressively constructs the Pareto front through a closed loop between front evolution and model updating.}
\label{fig:pipeline}
\end{figure}

We propose a manufacturability-constrained lattice optimization method for additive manufacturing and a progressive mechanism for Pareto-front construction. The overall framework is shown in Fig.~\ref{fig:pipeline}. It has two levels. At the first level, manufacturable lattice unit cells are generated for prescribed design conditions by inverse-homogenization topology optimization subject to manufacturing constraints. At the second level, a Pareto front is constructed through an iterative closed loop between front evolution and model updating, thereby characterizing performance trade-offs within the manufacturable region.

We first define the problem, then describe inverse-homogenization optimization of a unit cell under manufacturing constraints, and finally explain how the optimizer is used to progressively construct a manufacturable Pareto front.

\subsection{Problem Definition}
\label{subsec:problem_def}

Let the design domain of a periodic lattice unit cell be $\Omega\subset\mathbb{R}^3$, discretized into $N_e$ voxel elements. The design density field is $\rho^d=\{\rho^d_e\}_{e=1}^{N_e}$, where $\rho^d_e\in[0,1]$. To suppress checkerboard patterns, control the minimum feature size, and obtain clear solid--void boundaries, the design field is filtered and projected~\cite{sigmund200199} to obtain the physical density field used for physical analysis and manufacturability evaluation:
\begin{equation}
\rho^{\mathrm{phys}}=\mathcal{H}\big(\mathcal{F}(\rho^d)\big),
\end{equation}
where $\mathcal{F}(\cdot)$ and $\mathcal{H}(\cdot)$ denote density filtering and projection, respectively; their specific forms are given in Appendix~\ref{app:TO}. The physical density satisfies $\rho^{\mathrm{phys}}_e\in[\rho_{\min},1]$, where $\rho_{\min}$ is introduced to avoid numerical singularities. Values close to $1$ represent solid, whereas values close to $\rho_{\min}$ represent void.

Using $\rho^{\mathrm{phys}}$ as input, we define a multiphysics performance vector in terms of the effective Young's modulus $E^H$ and effective thermal conductivity $\kappa^H$:
\begin{equation}
\mathbf{F}(\rho^{\mathrm{phys}})=
[E^H(\rho^{\mathrm{phys}}),\kappa^H(\rho^{\mathrm{phys}})].
\end{equation}
The manufacturability vector is
\begin{equation}
\mathbf{M}(\rho^{\mathrm{phys}})=
\left[P_{\rm oh}(\rho^{\mathrm{phys}};\theta_{\rm oh}),
P_{\rm cav}(\rho^{\mathrm{phys}};r_{\rm ch})\right],
\end{equation}
where $P_{\rm oh}$ measures overhang risk and $\theta_{\rm oh}$ is the minimum self-support angle. The unified void-manufacturing penalty $P_{\rm cav}$ accounts for both enclosed-cavity risk and ineffective connectivity under a minimum powder-removal channel radius $r_{\rm ch}$.

The multiphysics optimization problem for a periodic lattice unit cell is therefore
{\small
\begin{equation}
\begin{aligned}
&\max_{\rho^d}\quad
\mathbf{F}\big(\rho^{\mathrm{phys}}(\rho^d)\big)
=\left[E^H\big(\rho^{\mathrm{phys}}(\rho^d)\big),
\kappa^H\big(\rho^{\mathrm{phys}}(\rho^d)\big)\right],\\
&\mathrm{s.t.}\quad
\frac{1}{N_e}\sum_{e=1}^{N_e}\rho^{\mathrm{phys}}_e(\rho^d)\le v_f,
\quad \rho^d_e\in[0,1],
\quad \rho^{\mathrm{phys}}(\rho^d)\in\mathcal{M},
\end{aligned}
\end{equation}
}
where $v_f$ is the target volume fraction evaluated from the physical density field, and $\mathcal{M}$ denotes the feasible region defined by the AM constraints. Unless noted otherwise, all effective properties, manufacturability measures, and volume fractions below are evaluated from $\rho^{\mathrm{phys}}$.

For two manufacturable candidates $s_a$ and $s_b$, $s_a$ dominates $s_b$ if it is no worse in every objective and strictly better in at least one. All candidates that are not dominated by another manufacturable design form the Pareto set, whose distribution in objective space is the Pareto front to be constructed.

For a prescribed design condition, we first solve for an individual candidate within the feasible region (Section~\ref{subsec:single_opt}) and then construct a nondominated set (Section~\ref{subsec:pareto}) to characterize the multiphysics trade-off boundary.

\subsection{Inverse-Homogenization Topology Optimization under Manufacturing Constraints}
\label{subsec:single_opt}

An individual lattice sample is denoted by
\(
s=\{\rho^d,c,d,\mathbf{F},\mathbf{M}\},
\)
where $\rho^d$ is the design density field and $\mathbf{F}$ and $\mathbf{M}$ are its performance and manufacturability vectors. The design condition is
\(
c=[w,v_f],
\)
where $w$ is the performance-weight vector and $v_f\in[0,1]$ is the target volume fraction. The manufacturing parameters are
\(
d=[\lambda_{\rm oh},\lambda_{\rm cav},\theta_{\rm oh},r_{\rm ch}],
\)
which contain the overhang- and cavity-penalty weights, the minimum self-support angle, and the minimum powder-removal channel radius, respectively. Under these conditions, unit-cell design is formulated as manufacturability-constrained inverse-homogenization topology optimization.

\subsubsection{Density Representation and Material Interpolation}

The periodic unit cell is discretized on a three-dimensional voxel grid. A continuous variable $\rho^d_e$ represents the material distribution in voxel $e$, and $\rho^{\mathrm{phys}}_e$ is used for physical analysis and manufacturability evaluation. The material properties are interpolated continuously from the physical density as
\begin{equation}
\mathbf{C}_e(\rho^{\mathrm{phys}}_e)
=(\rho^{\mathrm{phys}}_e)^p\mathbf{C}_0,
\qquad
\kappa_e(\rho^{\mathrm{phys}}_e)
=(\rho^{\mathrm{phys}}_e)^p\kappa_0,
\end{equation}
where $\mathbf{C}_0$ and $\kappa_0$ are the elasticity tensor and thermal conductivity of the solid material, and $p$ is the penalization exponent. This interpolation suppresses intermediate densities and encourages a binary solid--void distribution.

\subsubsection{Numerical Homogenization of Multiphysics Properties}

For a prescribed physical density field, linear elasticity and steady-state heat conduction are solved under periodic boundary conditions to compute the macroscopic effective properties.

\noindent\textbf{Elastic homogenization.}
For the $ij$-th unit macroscopic strain mode, the unit-cell equilibrium equation is
\begin{equation}
\nabla\cdot\left[
\mathbf{C}(\rho^{\mathrm{phys}}):
\left(\boldsymbol{\varepsilon}_{ij}^{0}
+\boldsymbol{\varepsilon}(\boldsymbol{\chi}^{ij})\right)
\right]=0,
\quad \mathrm{in}\ \Omega,
\end{equation}
where $\boldsymbol{\varepsilon}_{ij}^{0}$ is a unit macroscopic strain mode and $\boldsymbol{\chi}^{ij}$ is the periodic displacement fluctuation. Volume averaging gives the effective elasticity tensor
\begin{equation}
C_{ijkl}^{H}=\frac{1}{|\Omega|}\int_{\Omega}
\left(\boldsymbol{\varepsilon}_{ij}^{0}+\boldsymbol{\varepsilon}(\boldsymbol{\chi}^{ij})\right)
:\mathbf{C}(\rho^{\mathrm{phys}}):
\left(\boldsymbol{\varepsilon}_{kl}^{0}+\boldsymbol{\varepsilon}(\boldsymbol{\chi}^{kl})\right)
\dif\Omega.
\end{equation}
Mechanical measures such as the effective Young's modulus $E^H$ are extracted from $\mathbf{C}^{H}$.

\noindent\textbf{Thermal homogenization.}
Let $\mathbf{e}_i$ be a unit temperature-gradient direction and $\psi_i$ the corresponding periodic temperature fluctuation. The local steady-state conduction equation is
\begin{equation}
\nabla\cdot\left[
\kappa(\rho^{\mathrm{phys}})(\mathbf{e}_i+\nabla\psi_i)
\right]=0,
\quad \mathrm{in}\ \Omega.
\end{equation}
The effective conductivity tensor is then
\begin{equation}
\kappa_{ij}^{H}=\frac{1}{|\Omega|}\int_{\Omega}
\kappa(\rho^{\mathrm{phys}})
(\mathbf{e}_i+\nabla\psi_i)\cdot
(\mathbf{e}_j+\nabla\psi_j)
\dif\Omega.
\end{equation}
Homogenization thus maps $\rho^{\mathrm{phys}}$ to the multiphysics vector $\mathbf{F}(\rho^{\mathrm{phys}})$. To generate candidates with different preferences in a single optimization framework, we use the weighted scalar objective
\begin{equation}
J_{\mathrm{perf}}(\rho^{\mathrm{phys}})=
w_EE^{H}(\rho^{\mathrm{phys}})+w_\kappa\kappa^{H}(\rho^{\mathrm{phys}}),
\end{equation}
where $w_E$ and $w_\kappa$ are the elastic and thermal weights, respectively, and $w=[w_E,w_\kappa]$. Varying $w$ produces candidates with different performance preferences. Additional homogenization details are provided in Appendix~\ref{app:homogenization-theory}.

\subsubsection{Manufacturing-Constraint Modeling}
\begin{figure}[t]
\centering
\includegraphics[width=\linewidth]{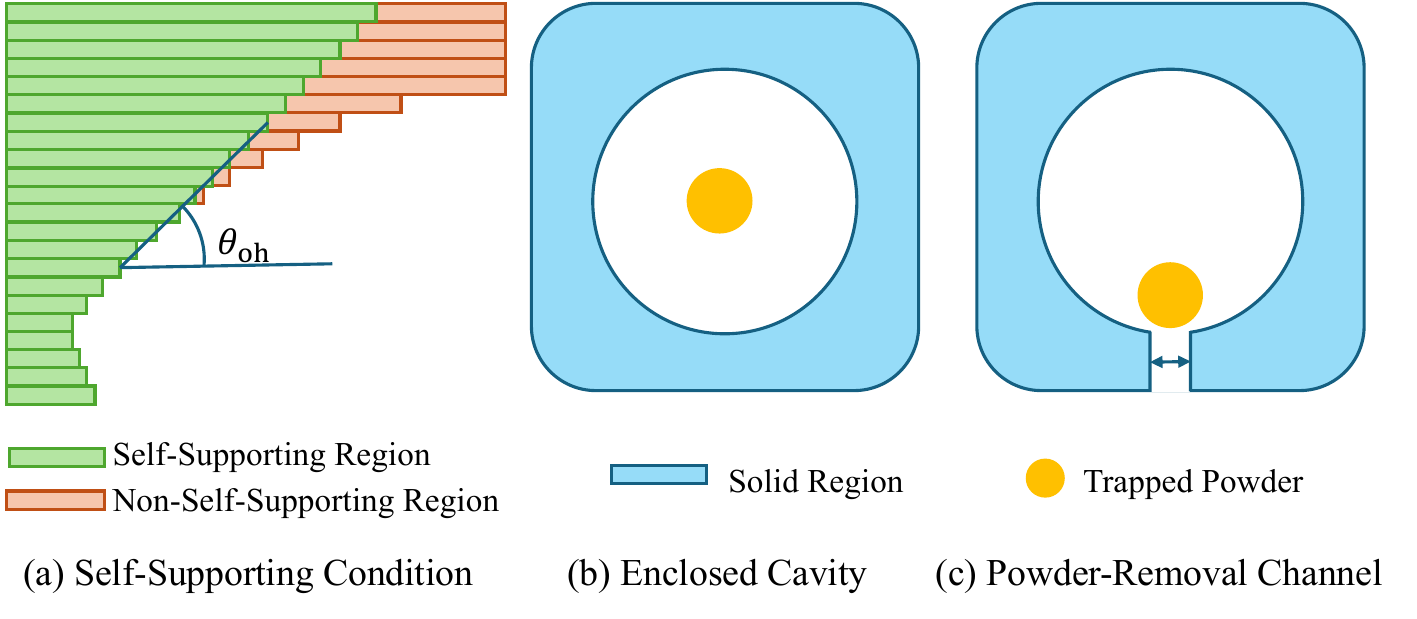}
\caption{\textbf{Additive-manufacturing constraints.} (a) Self-support: green denotes supported material and orange denotes a potential overhang. (b) Enclosed cavity: a white void is surrounded by solid and disconnected from the exterior. (c) Powder-removal channel: yellow denotes trapped powder caused by an undersized channel.}
\label{fig:printability}
\end{figure}

To suppress nonmanufacturable features during optimization, we formulate differentiable penalties on $\rho^{\mathrm{phys}}$ and incorporate them into the inverse-homogenization objective. The periodic unit cell is discretized into $n\times n\times n$ voxels and printed along $+z$. We consider three risks common to metal powder-bed fusion and related AM processes: local overhangs, enclosed cavities, and restricted powder-removal channels (Fig.~\ref{fig:printability}). The minimum-channel requirement is incorporated into the cavity-connectivity test rather than introduced as a separate penalty: a void path is considered effective only when it satisfies the prescribed channel scale.

\noindent\textbf{(1) Overhang constraint.}
In metal AM, structures are built layer by layer. If the local surface angle relative to the build platform falls below the admissible self-support angle, the surface may collapse or require additional support. Inspired by the layerwise self-support filter of~\cite{langelaar2016topology}, we construct a penalty on the continuous density field without directly modifying the field used for physical analysis.

Let $\theta_{\rm oh}$ be the minimum self-support angle. For isotropic voxels, the largest admissible lateral support distance between adjacent layers is
\begin{equation}
r_{\rm oh}=\left\lceil\tan\left(90^\circ-\theta_{\rm oh}\right)\right\rceil.
\end{equation}
For example, $\theta_{\rm oh}=45^\circ$ gives $r_{\rm oh}=1$, meaning that material in the current layer can be supported by material within a radius-one neighborhood in the layer below.

Let $S^k$ be the support field that can transfer support upward from layer $k$. Because the bottom layer contacts the build platform, it is initialized as
\begin{equation}
S^0_{i,j}=\rho^{\mathrm{phys}}_{i,j,0}.
\end{equation}
For layer $k$, the support field of the preceding layer is first dilated in the $xy$ plane to determine the support available to the current layer:
\begin{equation}
\bar S^k_{i,j}=\max_{(u,v)\in\mathcal N_{r_{\rm oh}}(i,j)}S^{k-1}_{u,v},
\end{equation}
where $\mathcal N_{r_{\rm oh}}(i,j)$ is a discrete neighborhood of radius $r_{\rm oh}$ centered at $(i,j)$. Periodic padding is used in $x$ and $y$ so that lateral support remains consistent after tiling.

If $\rho^{\mathrm{phys}}_{i,j,k}$ exceeds the support capacity $\bar S^k_{i,j}$, the excess is treated as potentially unsupported material. The overhang penalty is
\begin{equation}
P_{\rm oh}(\rho^{\mathrm{phys}})=
\frac{1}{n-1}\sum_{k=1}^{n-1}\frac{1}{n^2}\sum_{i,j}
\left[\rho^{\mathrm{phys}}_{i,j,k}-\bar S^k_{i,j}\right]_+,
\end{equation}
where $[x]_+=\max(x,0)$. Only supported material is allowed to propagate support upward:
\begin{equation}
S^k_{i,j}=\min\left(\rho^{\mathrm{phys}}_{i,j,k},\bar S^k_{i,j}\right).
\end{equation}
This recursion has the same physical interpretation as layerwise self-support filtering, but the support field is used only to construct $P_{\rm oh}$ and does not replace the density distribution used for homogenization.

\noindent\textbf{(2) Enclosed-cavity penalty based on a soft void indicator.}
An enclosed cavity is disconnected from the exterior and can trap powder. To construct a differentiable cavity penalty, we first define the soft void indicator
\begin{equation}
V(\rho^{\mathrm{phys}})=
\operatorname{sigmoid}\left(\beta_{\rm void}(\xi-\rho^{\mathrm{phys}})\right),
\end{equation}
where $\xi=0.5$ is the void threshold and $\beta_{\rm void}$ controls steepness. If $\rho^{\mathrm{phys}}<\xi$, then $V(\rho^{\mathrm{phys}})\approx1$ (void); if $\rho^{\mathrm{phys}}>\xi$, then $V(\rho^{\mathrm{phys}})\approx0$ (solid). This soft representation preserves a smooth density transition and makes the cavity penalty compatible with gradient-based optimization.

\begin{figure}[t]
\centering
\includegraphics[width=.8\linewidth]{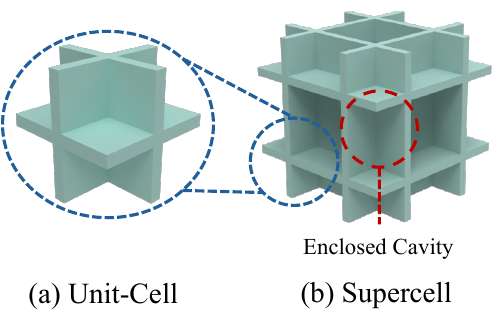}
\caption{\textbf{Difference between void connectivity in an isolated cell and in a periodic array.} (a) Connectivity at the unit-cell scale; (b) connectivity after periodic tiling into a supercell. A void that appears open in an isolated cell may become enclosed after tiling, so external reachability is evaluated on a supercell.}
\label{fig:padding}
\end{figure}

Starting from the exterior inlets, the externally reachable region is propagated through the soft void field. Let $B$ be the set of exterior seed voxels. The initial reachable field is
\begin{equation}
R^0=V(\rho^{\mathrm{phys}})\odot B,
\end{equation}
where $\odot$ denotes pointwise multiplication. A neighborhood-expansion operator then updates the field recursively:
\begin{equation}
\label{eq:cavity-expand}
R^{m+1}=\min\left(V(\rho^{\mathrm{phys}}),\mathcal P(R^m)\right),
\quad m=0,1,\ldots,K-1.
\end{equation}
The operator $\mathcal P(\cdot)$ expands the field over a local radius-one neighborhood, and $K$ is the number of propagation steps. Each update advances by one adjacent void voxel and remains bounded by $V(\rho^{\mathrm{phys}})$, so propagation cannot pass through solid material.

After $K$ steps, $R^K$ represents voids reachable from the exterior, whereas $V(\rho^{\mathrm{phys}})-R^K$ represents unreachable voids and hence potential enclosed cavities. The penalty is
\begin{equation}
P_{\rm cav}(\rho^{\mathrm{phys}})=
\frac{1}{|\Omega|}\sum_{\mathbf{x}\in\Omega}
\left[V(\rho^{\mathrm{phys}}(\mathbf{x}))-R^K(\mathbf{x})\right]_+.
\end{equation}

Connectivity at the boundary of a periodic unit cell does not necessarily reflect connectivity after tiling. As illustrated in Fig.~\ref{fig:padding}, a void may touch the boundary and appear open in an isolated cell, yet adjacent solid boundaries may jointly enclose it in a periodic array. We therefore expand the unit cell into a periodic supercell and propagate reachability from the outer supercell boundary, avoiding errors caused by truncation at a unit-cell boundary.

\begin{figure}[t]
\centering
\includegraphics[width=\linewidth]{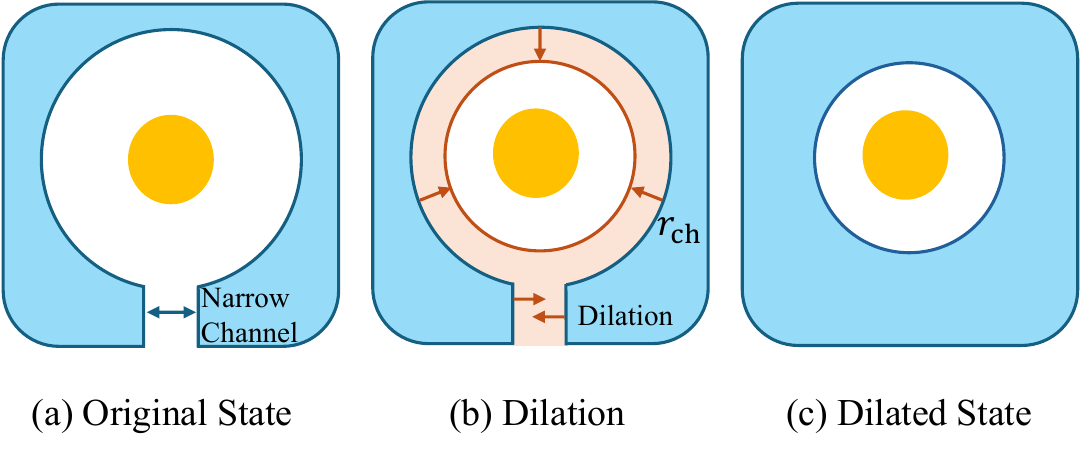}
\caption{\textbf{Minimum-channel control and cavity penalization based on void-channel-scale filtering.} (a) Original soft void field $V(\rho^{\rm phys})$; (b) morphological dilation of the soft solid by a radius of $r_{\rm ch}$, which blocks narrow channels; and (c) effective void field $V_{\rm ch}(\rho^{\rm phys})$ and the cavity penalty based on external reachability.}
\label{fig:channel_constraint}
\end{figure}

\noindent\textbf{(3) Minimum-channel control by void-channel-scale filtering.}
Some voids are connected to the exterior only through channels too narrow for effective powder removal. We therefore incorporate a minimum powder-removal channel requirement into the connectivity test so that only paths of sufficient width are treated as effective.

As shown in Fig.~\ref{fig:channel_constraint}, we define the soft solid field as $S(\rho^{\mathrm{phys}})=1-V(\rho^{\mathrm{phys}})$ and dilate it by $r_{\rm ch}$ voxels within each $xy$ cross section. This operation filters the void space at the prescribed channel scale: narrow passages that cannot accommodate the dilation radius are blocked by the dilated solid field and excluded from subsequent reachability propagation. Denoting the dilated soft solid field by $\widetilde S(\rho^{\mathrm{phys}})$, the effective void field is
\begin{equation}
V_{\rm ch}(\rho^{\mathrm{phys}})=
V(\rho^{\mathrm{phys}})\left(1-\widetilde S(\rho^{\mathrm{phys}})\right).
\end{equation}
External reachability is then propagated through $V_{\rm ch}(\rho^{\mathrm{phys}})$ according to Eq.~\ref{eq:cavity-expand}. The channel-aware cavity penalty is
\begin{equation}
P_{\rm cav}(\rho^{\mathrm{phys}};r_{\rm ch})=
\frac{1}{|\Omega|}\sum_{\mathbf{x}\in\Omega}
\left[V_{\rm ch}(\rho^{\mathrm{phys}}(\mathbf{x}))-R^K(\mathbf{x})\right]_+.
\label{eq:p-cav}
\end{equation}
Completely enclosed voids and voids connected only through undersized channels are thus penalized in a unified formulation.

Channel-scale filtering removes microvoids that cannot accommodate the minimum radius $r_{\rm ch}$ from $V_{\rm ch}$, so they do not directly contribute to Eq.~\ref{eq:p-cav}. Here $r_{\rm ch}$ denotes the smallest channel radius that is meaningful for powder removal. Smaller isolated microvoids neither provide useful powder-removal space nor substantially affect the effective properties and consequently have little incentive to form in the present optimization. We did not observe such cavities in the final structures, so they do not materially affect the manufacturability or performance conclusions.

In the implementation, neighborhood dilation and erosion are realized by pooling-based morphology retained in the automatic-differentiation graph. The sigmoid is differentiable, while $\min(\cdot)$, max pooling, and $[\cdot]_+$ are continuous and piecewise differentiable.
We use \texttt{torch.minimum} for the minimum operation and \texttt{ReLU} for the positive-part operation. At ties and switching points, the optimizer uses the subgradients selected by the framework.

\subsubsection{Topology-Optimization Procedure}

Combining multiphysics performance with the manufacturing penalties, we formulate the unit-cell problem for a prescribed condition $c$ as
{\small
\begin{equation}
\begin{aligned}
\max_{\rho^d}\quad
&J(\rho^{\mathrm{phys}},c)=
J_{\rm perf}(\rho^{\mathrm{phys}})
-\lambda_{\rm oh}P_{\rm oh}(\rho^{\mathrm{phys}})
-\lambda_{\rm cav}P_{\rm cav}(\rho^{\mathrm{phys}};r_{\rm ch})\\
\mathrm{s.t.}\quad&
\frac{1}{N_e}\sum_{e=1}^{N_e}\rho^{\mathrm{phys}}_e\le v_f,
\quad \rho^d_e\in[0,1],
\quad e=1,2,\ldots,N_e.
\end{aligned}
\end{equation}
}
Here $J_{\rm perf}$ is the weighted performance objective, $P_{\rm oh}$ is the overhang penalty, $P_{\rm cav}$ is the channel-aware void penalty, and $\lambda_{\rm oh}$ and $\lambda_{\rm cav}$ control their relative weights.

Given an initial design density $\rho^d_0$, the design variables are updated using the optimality-criteria (OC) method~\cite{berke1987structural}. Each iteration performs the following operations:
\begin{enumerate}[leftmargin=*]\setlength\itemsep{1mm}
    \item numerical homogenization to compute effective mechanical and thermal properties;
    \item evaluation of $P_{\rm oh}(\rho^{\mathrm{phys}})$ and $P_{\rm cav}(\rho^{\mathrm{phys}};r_{\rm ch})$;
    \item sensitivity analysis and a density update under the volume constraint; and
    \item filtering and projection to suppress checkerboards, control the minimum scale, and sharpen the solid--void boundary.
\end{enumerate}
Optimization terminates when the objective change falls below a threshold or the maximum number of iterations is reached. It outputs the candidate unit cell together with its performance and manufacturability measures.

This procedure generates one manufacturable candidate for a prescribed condition but does not characterize the global trade-off. We next use it as the inner solver in a progressive Pareto-front construction process.

\subsection{Progressive Pareto-Front Construction}
\label{subsec:pareto}

The high dimensionality, strong nonconvexity, and initialization sensitivity of inverse homogenization make it difficult to obtain a representative Pareto front through random initialization and repeated optimization alone. We therefore couple structural-prior learning with front construction in a closed loop between front evolution and model updating.

\subsubsection{Overview}

A single topology-optimization run yields only one candidate for a prescribed condition and initial density. Constructing a representative front requires repeated exploration of the joint space of design conditions and target properties, followed by feasibility and nondominance filtering. We treat this process as a progressive, data-driven construction process. Random exploration first produces an initial front and a set of high-quality samples. A target-performance-conditioned density generation network then learns a mapping from design conditions and target properties to topological priors. Latent representations of neighboring nondominated solutions are interpolated in front regions with sparse coverage, decoded into initial density fields, and verified through topology optimization. Accepted structures update both the front and the network, coordinating front evolution with model learning.

Let $\mathcal{D}^{(i)}$ be the high-quality sample set in iteration $i$ and $\mathcal{S}^{(i)}$ its Pareto set. Rather than approximating the entire front in one pass, the method adds solutions iteratively to improve representation quality and extend the trade-off boundary.

\subsubsection{Stage I: Random Exploration and Initial-Front Construction}

Stage I explores the design-condition space to cover the principal trade-off region and create the high-quality training set. For each sampled condition $c$, multiple random density fields are generated and optimized under the manufacturing constraints.

The optimized structures first undergo a manufacturability filter: candidates whose overhang, enclosed-cavity, or channel measures exceed their thresholds are removed. Nondominated filtering is then applied to the feasible set to obtain the Stage-I Pareto set. Because a strict Pareto set may be too small for network training, near-Pareto samples with favorable nondominated ranks, short distances to the front, and good manufacturability are also retained to increase both sample count and topological diversity.

The resulting high-quality set is
\(
\mathcal{D}^{(0)}=\{s_i\}_{i=1}^{N},
\)
which is used to train the target-performance-conditioned density generation network.

\subsubsection{Target-Performance-Conditioned Density Generation Network}

To improve the efficiency of front construction, we introduce a density generation network $G_\theta$ conditioned on target performance. Unlike a network driven only by design parameters, $G_\theta$ explicitly takes a target-performance vector $\mathbf f^\star$ that specifies the desired location in property space. It thereby learns a high-quality topological prior jointly constrained by the design condition and the target properties. The encoder--decoder architecture enables interpolation within the learned latent space. A convex combination of neighboring nondominated latent representations is decoded into an initial design field that tends to lie near the performance trade-off boundary. The network output serves only as the initialization; the final structure is determined by constrained topology optimization.
\begin{figure}[t]
\centering
\includegraphics[width=\linewidth]{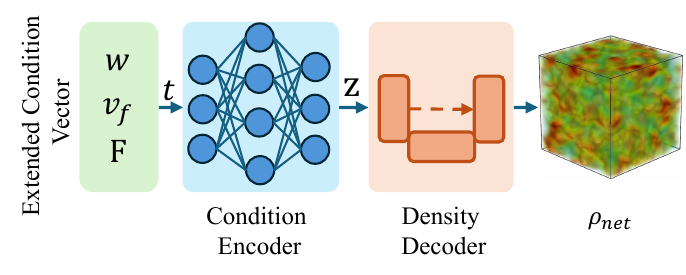}
\caption{\textbf{Target-performance-conditioned density generation network.} The extended condition vector is encoded into a latent representation and decoded into an initial density field for topology optimization.}
\label{fig:network-pipeline}
\end{figure}

As shown in Fig.~\ref{fig:network-pipeline}, the network takes the extended condition vector $\mathbf t=[\mathbf w,v_f,\mathbf f^\star]$,
where $\mathbf w$ specifies the performance preference, $v_f$ specifies the material usage, and $\mathbf f^\star$ specifies the target location in performance space. During training, sample $i$ uses $\mathbf f_i^\star=\mathbf F(\rho_i^{\mathrm{phys},\ast})$; during front expansion, $\mathbf f^\star$ is generated from a front region to be filled. The network consists of a condition encoder $E$ and a three-dimensional decoder $D$:
\begin{equation}
z=E(\mathbf t),\qquad
\rho_{\mathrm{net}}=D(z)=G_\theta(\mathbf t).
\end{equation}
The decoder is implemented using a three-dimensional U-Net, in which multiscale convolutions and skip connections are employed to capture both local and global geometric features. A final sigmoid followed by affine rescaling maps the output to $[\rho_{\min},1]$:
\begin{equation}
\rho_{\mathrm{net}}=\rho_{\mathrm{min}}+(1-\rho_{\mathrm{min}})\operatorname{sigmoid}(o_\theta),
\end{equation}
where $o_\theta$ denotes the output of the final network layer. The resulting density field serves as the initial design density for topology optimization. The subsequent density filtering, projection, and construction of the physical density field used in finite-element analysis are described in Appendix~\ref{app:TO}. The network architecture and hyperparameters are detailed in Appendix~\ref{app:network_para}.

The network is trained on $\mathcal{D}^{(i)}$ using
{\small
\begin{equation}
\begin{aligned}
&\mathcal{L}=\mathcal{L}_{\mathrm{rec}}
+a\mathcal{L}_{\mathrm{vol}}
+b\mathcal{L}_{\mathrm{smooth}},\\
&\mathcal{L}_{\mathrm{rec}}
=\left\|G_\theta(\mathbf t_i)-\rho_i^\ast\right\|_2^2,\\
&\mathcal{L}_{\mathrm{vol}}
=\frac{1}{N}\sum_{i=1}^{N}
\left|\frac{1}{N_e}\sum_{e=1}^{N_e}G_\theta(\mathbf t_i)_e-v_{f,i}\right|,\\
&\mathcal{L}_{\mathrm{smooth}}
=\left\|\nabla G_\theta(\mathbf t_i)\right\|_2^2.
\end{aligned}
\end{equation}
}
The reconstruction loss encourages the output to match high-quality priors, the volume loss enforces consistency with $v_f$, and the smoothness loss suppresses high-frequency noise that would hinder subsequent optimization.

\subsubsection{Stage II: Network-Guided Iterative Front Construction}

Stage II improves the nondominated set and extends the trade-off boundary from the initial front. New target locations are interpolated within front regions, the network predicts the associated initial design densities, and manufacturability-constrained topology optimization validates the resulting candidates. Each iteration contains four steps, as shown in Stage II of Fig.~\ref{fig:pipeline}.

\noindent\textbf{(1) Front-region identification and adaptive interpolation.}

Let $\mathcal{D}^{(i)}$ be the current high-quality set and $\mathcal{S}^{(i)}$ its nondominated subset. To concentrate the optimization budget on underrepresented front regions, we identify intervals directly in performance space. The performance vector of each nondominated sample is first normalized:
\begin{equation}
\widehat{\mathbf F}(s)=
\left[
\frac{F_1(s)-F_1^{\min}}{F_1^{\max}-F_1^{\min}+\epsilon},
\ldots,
\frac{F_m(s)-F_m^{\min}}{F_m^{\max}-F_m^{\min}+\epsilon}
\right],
\end{equation}
where $m$ is the number of objectives, $F_j^{\min}$ and $F_j^{\max}$ are the minimum and maximum values of objective $j$ on the current front, and $\epsilon=10^{-8}$ prevents division by zero. This $\widehat{\mathbf F}$ is used only to identify front neighborhoods. The target-performance vector supplied to the network is denoted by $\mathbf f(s)$ and uses the same fixed normalization as in training.

In the two-objective case, let $w(s)$ denote the scalar ordering
coordinate associated with solution $s\in\mathcal{S}^{(i)}$.
For a target coordinate $w^\ast$, define
\begin{equation}
\begin{aligned}
s_l &= \operatorname*{arg\,max}_{s\in\mathcal{S}^{(i)}:\,w(s)<w^\ast} w(s),\\
s_u &= \operatorname*{arg\,min}_{s\in\mathcal{S}^{(i)}:\,w(s)>w^\ast} w(s).
\end{aligned}
\end{equation}
Thus, $\mathcal{N}(w^\ast)=\{s_l,s_u\}$.
Their latent representations are linearly interpolated:
\begin{equation}
z^\ast=(1-\alpha)z_l+\alpha z_u,
\qquad z_q=E(\mathbf t_q),
\qquad \alpha\sim U(0,1),
\end{equation}
where $\mathbf t_q=[\mathbf w_q,v_{f,q},\mathbf f_q]$ is the extended condition vector of neighbor $q$.

For three objectives, the front forms a nondominated surface rather than a polyline. We therefore define a local neighborhood in normalized performance space. For a front sample $s$, its $K$ nearest neighbors are
\begin{equation}
\mathcal{N}(s)=
\operatorname{KNN}\left(
\widehat{\mathbf F}(s),
\{\widehat{\mathbf F}(q)\mid q\in\mathcal{S}^{(i)},q\ne s\}
\right).
\end{equation}
For each $s$, convex-combination weights are sampled uniformly as $\boldsymbol{\eta}\sim\mathrm{Dir}(1,\ldots,1)$, and a latent representation is formed by interpolating $z_s$ with those of its neighbors:
\begin{equation}
z^\ast=\sum_{q\in\{s\}\cup\mathcal N(s)}\eta_q z_q,
\qquad z_q=E(\mathbf t_q).
\end{equation}

Each iteration generates $B$ interpolated representations $\{z_k^{\star(i+1)}\}_{k=1}^{B}$. The two-objective case reduces to linear interpolation between adjacent samples, whereas the three-objective case uses a convex combination within a local neighborhood.

\noindent\textbf{(2) Latent decoding and initial-density generation.}
Each interpolated representation is decoded as
\begin{equation}
\rho_k^{\ast(i+1)}=D\left(z_k^{\ast(i+1)}\right).
\end{equation}
Because the latent space is learned from high-quality solutions, the decoded field tends to lie close to the performance trade-off boundary, and optimization initialized from it is more likely to converge to a high-quality manufacturable solution than optimization initialized from a random or directly interpolated density field.

\noindent\textbf{(3) Topology optimization and front updating.}
Each $\rho_k^{\ast(i+1)}$ initializes manufacturability-constrained topology optimization. The performance and manufacturability of each new candidate are evaluated, accepted candidates are merged with the existing set, and nondominated filtering produces $\mathcal{D}^{(i+1)}$ and $\mathcal{S}^{(i+1)}$.

\noindent\textbf{(4) Data-set updating and network retraining.}
The updated $\mathcal{D}^{(i+1)}$ is used to train or fine-tune the network for the next iteration. New samples make the learned latent space more representative of the current front and enable later interpolants to approach the true trade-off boundary more closely, thereby coordinating front evolution with model learning.

\section{Experiments and Results}

\subsection{Experimental Setup}
\label{subsec:experimental_setup}

\paragraph{Optimization and training parameters.}
All experiments use three-dimensional periodic lattice unit cells. We use the resolution $r$ to denote the number of voxels along each coordinate direction, so a unit cell at resolution $r$ is discretized into an $r\times r\times r$ grid with $N_e=r^3$ voxel elements. Stage-I topology optimization uses $r=64$, whereas the network outputs densities at $r=32$. The minimum density is $\rho_{\min}=10^{-5}$, the material-interpolation exponent is $p=3$, and the maximum number of optimization iterations is $I_{\max}=2000$. The build direction is $+z$, the minimum self-support angle is $\theta_{\rm oh}=45^\circ$, the minimum powder-removal channel radius is $r_{\rm ch}=3$ voxels, the number of cavity-reachability propagation steps is $K=128$, and the penalty weights are $\lambda_{\rm oh}=\lambda_{\rm cav}=1$. The network loss weights are $a=1.0$ and $b=10^{-3}$. All experiments were performed on a server with eight NVIDIA RTX 3090 GPUs.

\paragraph{Additive-manufacturing parameters.}
Metal specimens were fabricated from AlSi10Mg using an EOS M290 laser powder-bed fusion system. The process parameters were a laser power of $370$~W, a scan speed of $1300$~mm/s, a hatch spacing of $0.19$~mm, and a layer thickness of $0.03$~mm. These settings establish the process context for our manufacturability analysis and inform the overhang, powder-channel, and feature-scale constraints. The soft void indicator uses the threshold $\xi=0.5$ and steepness $\beta_{\rm void}=20$.

\paragraph{Manufacturability evaluation.}
For quantitative evaluation, manufacturability is assessed on a binary geometry rather than directly on the continuous physical density. We threshold at $\xi=0.5$: $\rho^{\mathrm{phys}}(\mathbf{x})\ge\xi$ defines the solid region $\Omega_s$, and the remainder defines the void region $\Omega_v$.

Connectivity checks performed only within an isolated periodic cell are susceptible to boundary-truncation errors. We therefore tile the cell into a $2\times2\times2$ array, denoted by $\Omega_s^{2\times2\times2}$, and evaluate manufacturability on this supercell. This size is sufficient to cover every boundary-truncation pattern produced by an arbitrary translation of a one-unit-cell-sized crop. Let the original unit-cell side length be $L$. Any axis-aligned crop of the same size can be written as $\Omega_{\boldsymbol\delta}=\boldsymbol\delta+[0,L]^3$, where $\boldsymbol\delta\in[0,L)^3$. In each coordinate direction, $[\delta_j,\delta_j+L]\subseteq[0,2L]$, so the crop intersects at most two adjacent cells. In three dimensions, every unit-cell-scale truncation, including configurations crossing faces, edges, and corners, is therefore contained in a $2\times2\times2$ supercell.

Consequently, the $2\times2\times2$ array is the smallest sufficient supercell for eliminating unit-cell boundary-truncation errors. A larger supercell merely repeats existing local periodic configurations without introducing a new crop pattern, while increasing the cost of morphological propagation and reachability analysis.

\subsection{Evaluation Metrics}

We evaluate three aspects of the method: (1) single-sample optimization, (2) characterization of multiphysics trade-offs, and (3) efficiency of progressive Pareto-front construction.

\paragraph{Optimization success rate.}
We first test convergence on the continuous physical density and then evaluate global connectivity on the thresholded solid geometry. Let $J^{(t)}$ be the objective at iteration $t$ and $I_{\max}$ the maximum number of iterations. A sample satisfies the convergence criterion if there exists a $t\le I_{\max}$ such that
\begin{equation}
\max_{\tau\in\{t-2,t-1,t\}}
\left|J^{(\tau)}-J^{(\tau-1)}\right|<5\times10^{-4}.
\end{equation}
We then set $C_{\rm conv}=1$; otherwise, $C_{\rm conv}=0$. For a converged sample, $\rho^{\mathrm{phys}}$ is thresholded and the connectivity of its solid network is evaluated.

Let $N_{\rm cc}$ be the number of connected components in $\Omega_s^{2\times2\times2}$. The periodically tiled solid is regarded as globally connected if $N_{\rm cc}=1$. The optimization-success indicator is
$S_{\rm opt}=\mathbb{I}[C_{\rm conv}=1\land N_{\rm cc}=1]$, and the optimization success rate $R_{\rm opt}$ is the proportion of all samples for which $S_{\rm opt}=1$. This metric measures the ability of topology optimization to produce a valid lattice but does not by itself measure manufacturability.

\paragraph{Effective-property metrics.}
For an optimized density, periodic homogenization provides the effective elasticity tensor $\mathbf{C}^H$ and thermal-conductivity tensor $\boldsymbol\kappa^H$, from which we extract the effective Young's modulus $E^H$, shear modulus $G^H$, and thermal conductivity $\kappa^H$. These values are normalized by the base-material properties $E_0$, $G_0$, and $\kappa_0$ as $\bar E=E^H/E_0$, $\bar G=G^H/G_0$, and $\bar\kappa=\kappa^H/\kappa_0$. Pareto-front analysis maximizes $\bar E$ and $\bar\kappa$. Dominance and nondominance follow the definitions in Section~\ref{subsec:problem_def}.

\paragraph{Manufacturability metrics.}
Let the subset of solid voxels satisfying the self-support criterion be
{\small
\[
\Omega_{\rm oh}=
\left\{
\mathbf{x}\in\Omega_s^{2\times2\times2}:
\mathbf{x}\ \text{satisfies the }\theta_{\rm oh}\text{ self-support criterion for a }+z\text{ build}
\right\}.
\]
}
The self-support score is
\begin{equation}
S_{\rm oh}=\frac{|\Omega_{\rm oh}|}{|\Omega_s^{2\times2\times2}|}.
\end{equation}
A larger $S_{\rm oh}$ indicates a lower overhang risk.

To assess cavity connectivity, we dilate $\Omega_s^{2\times2\times2}$ by radius $r_{\rm ch}$ to block narrow channels and denote the resulting ineffective void region by $\Omega_v^{\rm cav}$. The cavity score is
\begin{equation}
S_{\rm cav}=\frac{|\Omega_v^{\rm cav}|}{|\Omega_s^{2\times2\times2}|}.
\end{equation}
$S_{\rm cav}=0$ means that every void has an effective powder-removal path; larger values indicate a greater risk of trapped powder.

A structure is classified as manufacturable when $S_{\rm oh}\ge0.95$ and $S_{\rm cav}\le0.05$. These thresholds reflect practical experience with three-dimensional metal printing. The manufacturable fraction $R_{\rm feas}$ is the proportion of successfully optimized samples that satisfy both requirements. Thus, $S_{\rm opt}$ measures whether optimization produces a valid lattice, whereas $R_{\rm feas}$ further measures AM feasibility.

\paragraph{Pareto-front quality.}
We use hypervolume (HV) and the coverage metric (C-metric) to evaluate front quality~\cite{ishibuchi2019comparison}. Let $P$ be a filtered nondominated set and $\mathbf f(s)$ the objective vector of solution $s$, with every objective expressed as a maximization.

Hypervolume measures the objective-space region dominated by $P$ relative to a reference point and reflects both convergence and coverage. For a reference point $\mathbf r$ that is worse than all compared solutions in every objective,
{\small
\begin{equation}
HV(P)=\lambda_m\left(
\bigcup_{\mathbf f(s)\in P}
[r_1,f_1(s)]\times\cdots\times[r_m,f_m(s)]
\right),
\end{equation}
}
where $\lambda_m(\cdot)$ is the $m$-dimensional Lebesgue measure. A larger $HV$ indicates a better front.

Coverage compares the relative dominance of two fronts. For nondominated sets $A$ and $B$,
\begin{equation}
C(A,B)=
\frac{
\left|\{\mathbf b\in B\mid\exists\mathbf a\in A,\ \mathbf a\succeq\mathbf b\}\right|
}{|B|}.
\end{equation}
Thus $C(A,B)$ is the fraction of $B$ covered by $A$.

\subsection{Multiphysics Pareto Fronts}

\begin{figure}[t]
\centering
\includegraphics[width=\linewidth]{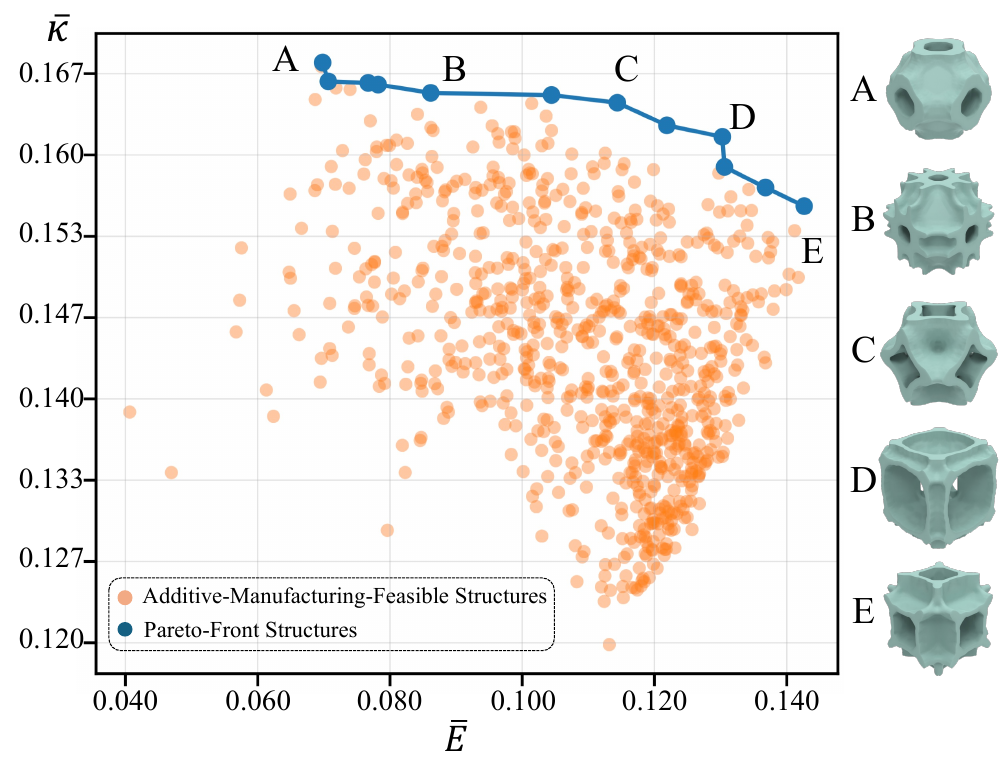}
\caption{\textbf{Pareto front for two-objective stiffness--conductivity optimization.} The horizontal and vertical axes show normalized effective stiffness $\bar E$ and thermal conductivity $\bar\kappa$, respectively. Orange points are candidate lattices satisfying the AM requirements, and blue points are nondominated solutions. Representative structures on the Pareto front, shown on the right, illustrate topology changes under different stiffness and conductivity preferences.}
\label{fig:pareto_2obj}
\vspace{-3mm}
\end{figure}

We first analyze the distribution of lattice structures under multiple physical objectives. All candidates are filtered using the manufacturability criteria in Section~\ref{subsec:experimental_setup}; only structures that satisfy the overhang, cavity, powder-channel, and connectivity requirements are retained. The two- and three-objective experiments fix $v_f=25\%$ to eliminate variation in material usage. A subsequent experiment varies $v_f$ to study its effect.

\paragraph{Two-objective stiffness--conductivity Pareto front.}
With $v_f=25\%$, different preferences are specified by the ratio $w_E/w_\kappa$:
\begin{equation}
J_{\rm perf}(\rho)=w_E\bar E(\rho)+w_\kappa\bar\kappa(\rho),
\qquad w_E+w_\kappa=1.
\end{equation}
We sample $w_E\in[0,1]$ and set $w_\kappa=1-w_E$.

Across 1000 initializations, the optimization success rate is $R_{\rm opt}=78.30\%$ and the manufacturable fraction is $R_{\rm feas}=99.11\%$, yielding 776 successfully optimized and manufacturable structures. Figure~\ref{fig:pareto_2obj} plots their distribution in the $\bar E$--$\bar\kappa$ plane. Twelve structures lie on a clear nondominated boundary, confirming a pronounced trade-off between stiffness and thermal conductivity.

The topologies reflect the corresponding preferences. A larger $w_E$ strengthens load-bearing paths and increases $\bar E$; a larger $w_\kappa$ produces more continuous heat-conduction paths and increases $\bar\kappa$; intermediate weights yield compromise structures. The method therefore generates manufacturable lattices with different functional preferences and effectively characterizes their Pareto boundary.

\begin{figure}[t]
\centering
\includegraphics[width=\linewidth]{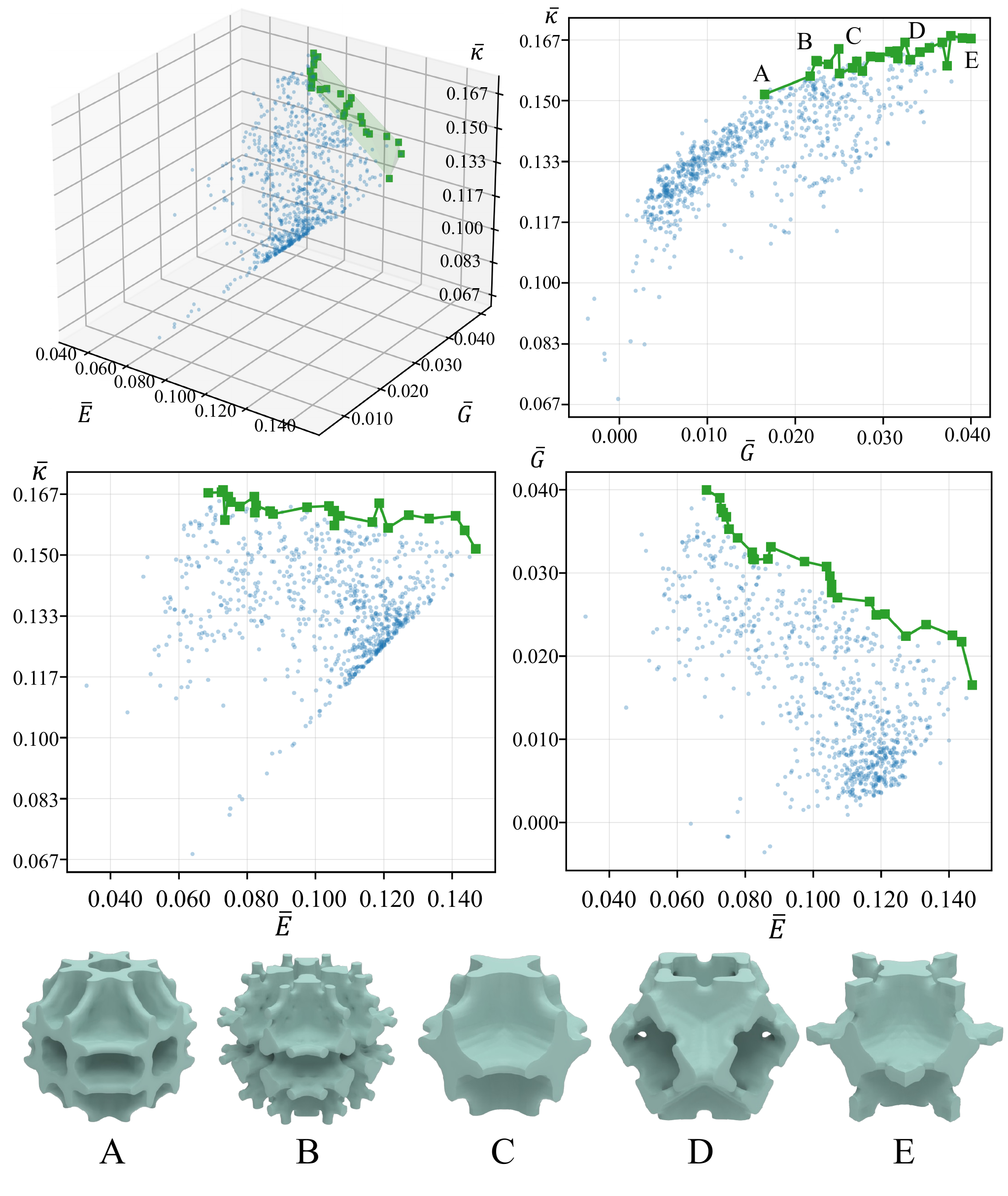}
\caption{\textbf{Three-objective optimization of stiffness, thermal conductivity, and shear modulus.} Candidates occupy the objective space defined by $\bar E$, $\bar\kappa$, and $\bar G$; green points are nondominated. The lower and right panels show projections onto the three pairwise objective planes, and representative Pareto structures appear below.}
\label{fig:pareto_3obj}
\vspace{-3mm}
\end{figure}

\paragraph{Three-objective stiffness--conductivity--shear Pareto front.}
We next include effective shear modulus to test the framework in a higher-dimensional objective space. The performance vector becomes
\begin{equation}
\mathbf F(\rho)=
\left[\bar E(\rho),\bar\kappa(\rho),\bar G(\rho)\right],
\end{equation}
where $\bar G=G^H/G_0$. The scalarized objective is
\begin{equation}
J_{\rm perf}(\rho)=
w_E\bar E(\rho)+w_\kappa\bar\kappa(\rho)+w_G\bar G(\rho),
\qquad w_E+w_\kappa+w_G=1.
\end{equation}
We fix $v_f=25\%$ and sample the three-objective weight simplex.

Figure~\ref{fig:pareto_3obj} shows the resulting three-objective distribution. With shear modulus included, the candidates form a nondominated surface with more complex competition among properties. Some structures balance stiffness and thermal conductivity well but have lower shear resistance, whereas structures with high shear modulus sacrifice axial stiffness or thermal conductivity. All three two-dimensional projections are well dispersed, showing that the method can construct a representative nondominated set in a higher-dimensional objective space.

Under the tested volume fraction, manufacturing parameters, and weight-sampling range, the projections show neither a pronounced concave region nor an isolated nondominated branch. Weighted scalarization therefore provides reasonably complete coverage in this experiment. This observation is specific to the tested setting and does not provide a theoretical guarantee of Pareto-front convexity or completeness under weighted scalarization.

\paragraph{Three-way trade-off among stiffness, conductivity, and lightweight design.}
\begin{figure}[t]
\centering
\includegraphics[width=\linewidth]{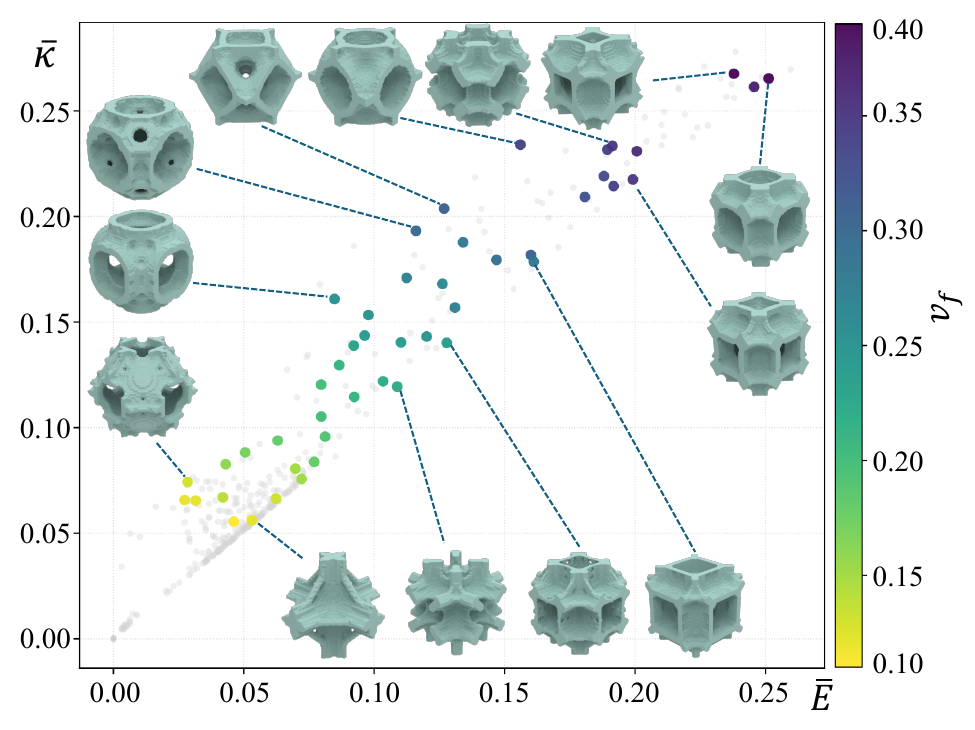}
\caption{\textbf{Trade-off among stiffness, thermal conductivity, and volume fraction.} Color denotes target volume fraction $v_f$, and points show optimized lattices in the $\bar E$--$\bar\kappa$ plane. Representative structures surround the plot. Increasing $v_f$ strengthens load-bearing and heat-conduction paths but compresses void space and may restrict powder removal.}
\label{fig:vf-performance}
\vspace{-4mm}
\end{figure}

Volume fraction affects not only material usage but also the mechanical performance, thermal performance, and manufacturability of a lattice. Unlike the preceding two-objective experiment at fixed volume fraction, this experiment samples combinations of performance weights and target volume fractions and solves a separate manufacturability-constrained inverse-homogenization problem for each combination. The resulting manufacturable structures are then filtered globally with three objectives: maximize normalized stiffness $\bar E$, maximize normalized thermal conductivity $\bar\kappa$, and minimize $v_f$, where $v_f\in[0.1,0.4]$.

As Fig.~\ref{fig:vf-performance} shows, both $\bar E$ and $\bar\kappa$ generally increase with $v_f$ because additional solid material produces more continuous load-bearing and heat-conduction paths. A larger volume fraction, however, consumes more material, compresses the void space, and increases the local risk of restricted powder removal. The low-$v_f$ region is lightweight but has limited stiffness and thermal conductivity. As $v_f$ increases, the skeleton becomes more continuous and both properties improve substantially. At high $v_f$, further performance gains are increasingly constrained by lightweight-design requirements and manufacturability. Thus, a larger $v_f$ is not unconditionally preferable; volume fraction, stiffness, and conductivity form a genuine three-way trade-off.

\subsection{Network-Guided Iterative Pareto-Front Construction}

To assess the target-performance-conditioned density network and the iterative expansion strategy, we compare three initialization methods: \textbf{random initialization}, \textbf{$K$-nearest-neighbor interpolation}, and \textbf{network initialization}. All three use the same volume fraction, weight range, manufacturing parameters, optimizer, and maximum iteration count as the two-objective stiffness--conductivity experiment. Each method receives a total budget of 1000 topology-optimization runs.

\noindent\textbf{(1) Random initialization.}
For each design condition, inverse-homogenization topology optimization under the same manufacturing constraints is initialized with a random density field. This strategy uses no prior sample information and provides the baseline for unguided front construction. It performs 1000 optimization runs.

\noindent\textbf{(2) $K$-nearest-neighbor interpolation.}
This strategy constructs an initial field from previously optimized samples. Let sample $j$ have condition $c_j=[\mathbf w_j,v_{f,j}]$ and optimized density $\rho_j$. For a target condition $c^\ast=[\mathbf w^\ast,v_f^\ast]$, distances are computed in normalized condition space and the nearest $K$ samples form its neighborhood. Because $v_f$ is fixed in the two-objective experiment, the condition space is determined primarily by $w_E$. We use $K=2$ and select the existing samples immediately to the left and right of the target weight. Their optimized densities are linearly interpolated according to the relative target weight. After resolution matching, periodic smoothing, and volume-fraction correction, the interpolated field initializes topology optimization. This method also performs 1000 optimization runs.

\begin{figure}[t]
\centering
\includegraphics[width=\linewidth]{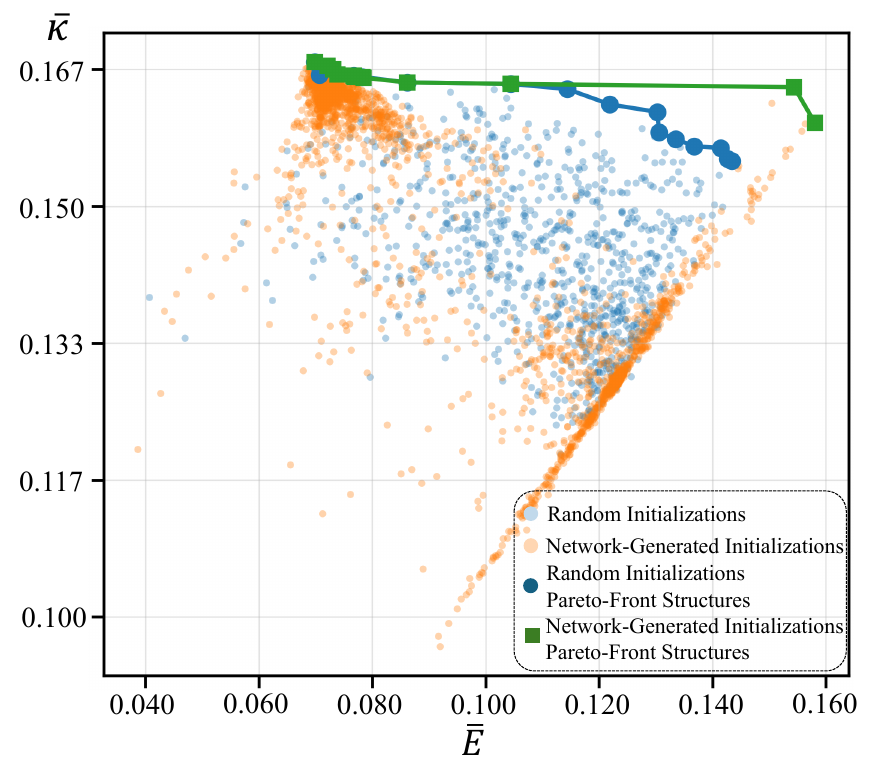}
\caption{\textbf{Pareto-front construction with random and network initialization.} Points denote candidates generated by the two strategies, and polylines denote their nondominated fronts. Under the same optimization budget, network initialization produces a higher-quality front.}
\label{fig:random-vs-network-pareto}
\end{figure}

\noindent\textbf{(3) Network-guided iterative expansion.}
For each selected Pareto-front region, latent representations of neighboring nondominated solutions are convexly interpolated and decoded into initial density fields. Manufacturability-constrained topology optimization then refines the fields. We perform 200 optimization runs per iteration for five iterations.

Figure~\ref{fig:random-vs-network-pareto} compares the objective-space distributions obtained by random and network initialization. Random initialization produces manufacturable structures but covers a limited performance range and is constrained by local optima. Under the same budget, network initialization shifts the front outward: it achieves higher conductivity while retaining high stiffness, and vice versa. The network therefore supplies topological priors that lie closer to a high-quality solution region.

\begin{table}[t]
\centering
\caption{\textbf{Optimization efficiency and manufacturability under different initialization strategies}}
\label{tab:init_feb}
\scalebox{0.70}{
\begin{tabular}{c|rrcccc}
\hline
Initialization
& $\overline I\downarrow$
& $R_{\rm opt}\uparrow$
& $\overline S_{\rm oh}\uparrow$
& $\overline S_{\rm cav}\downarrow$
& $R_{\rm feas}\uparrow$
& Manufacturable samples \\
\hline
Random & 1516 & 78.30\% & 0.96 & \bfseries{0.03} & \bfseries{99.11\%} & 776 \\
KNN & 881 & 86.40\% & 0.85 & 0.04 & 75.58\% & 653 \\
Network & \bfseries{484} & \bfseries{92.60\%} & \bfseries{0.96} & 0.04 & 98.92\% & \bfseries{916} \\
\hline
\end{tabular}
}
\end{table}

\paragraph{Efficiency and feasibility.}
Table~\ref{tab:init_feb} compares the three strategies. Random initialization requires 1516 iterations on average and achieves a $78.30\%$ success rate; 776 of its 783 successful samples are manufacturable. Nearest-neighbor interpolation reduces the mean iteration count to 881 and raises the success rate to $86.40\%$, but its mean self-support score drops to $0.85$ and only 653 samples are manufacturable. Direct density interpolation therefore accelerates convergence at the cost of additional manufacturing risks. Network initialization requires only 484 iterations on average and achieves a $92.60\%$ success rate; 916 of its 926 successful samples are manufacturable ($R_{\rm feas}=98.92\%$). Although its $R_{\rm feas}$ is slightly lower than that of random initialization, its much larger set of successful samples yields substantially more manufacturable samples.

\begin{table}[t]
\centering
\caption{\textbf{Pareto-front quality under different initialization strategies}}
\label{tab:init_quality_compare}
\scalebox{0.91}{
\begin{tabular}{c|ccc}
\hline
Initialization
& $HV(P)\uparrow$
& $C(A,\mathrm{Rand})\uparrow$
& $C(\mathrm{Rand},A)\downarrow$\\
\hline
Random & 0.0675 & 1.0000 & 1.0000 \\
KNN & 0.0620 & 0.5556 & 0.5714 \\
Network & \bfseries{0.0787} & \bfseries{1.0000} & \bfseries{0.4167} \\
\hline
\end{tabular}
}
\end{table}

\paragraph{Pareto-front quality.}
Table~\ref{tab:init_quality_compare} compares front quality. Random initialization achieves $HV=0.0675$, nearest-neighbor interpolation $HV=0.0620$, and network initialization $HV=0.0787$. The network therefore covers a larger objective-space region. Let $A$ denote the network-initialized front. Then $C(A,\mathrm{Rand})=1.0$, so all random-front solutions are covered by the network front, whereas $C(\mathrm{Rand},A)=0.4167$, so most network-front solutions are not covered by the random front. Network initialization shifts the Pareto front toward a better region overall.

Nearest-neighbor interpolation has the lowest hypervolume because linearly mixing densities can create disconnected or diffuse regions, placing the resulting initialization far from the performance boundary. By contrast, latent-space interpolation produces fields that more naturally approach the front and therefore converge more readily to high-quality solutions.

\paragraph{Computational cost.}
At resolution $64^3$, one topology-optimization run takes approximately 16 minutes, whereas training the network on 1000 samples takes less than 10 minutes. Network training is therefore negligible relative to repeated topology optimization, and the generated initial fields substantially reduce the number of subsequent optimization iterations.

\paragraph{Iterative front expansion.}
To examine convergence of the network-guided densification process, we record the hypervolume after each iteration. Random initialization yields an initial front with $HV=0.0675$. After network initialization, the first four iterations increase $HV$ to $0.0697$, $0.0732$, $0.0764$, and $0.0787$, respectively. Thus, the network-generated fields continually add effective front solutions and progressively move the front outward. The fifth iteration adds no solution that changes the current nondominated set, so the hypervolume remains unchanged. To assess the marginal benefit of continuing beyond 5 iterations, we run an additional sixth iteration from the iteration-5 front with three independent random seeds. The resulting hypervolumes are $0.0787$, $0.0787$, and $0.0823$, corresponding to a maximum change of $4.57\%$ relative to the iteration 5. Additional iterations therefore show diminishing but nonzero returns under the tested random seeds.

\begin{figure}[t]
\centering
\includegraphics[width=\linewidth]{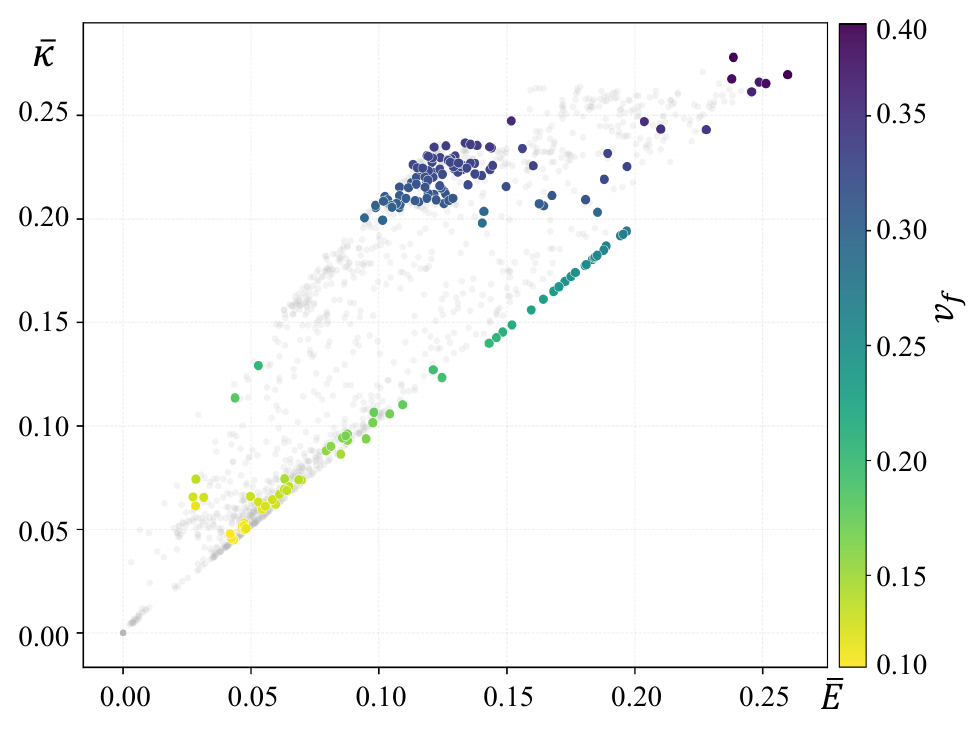}
\caption{\textbf{Three-objective nondominated set from joint training across volume fractions.} Gray points are the initial manufacturable samples obtained by random initialization, and colored points are additional samples generated by the jointly trained conditional network and refined by manufacturability-constrained topology optimization. Color denotes target volume fraction $v_f$. Global filtering uses normalized stiffness $\bar E$, normalized conductivity $\bar\kappa$, and material usage as the three objectives.}
\label{fig:vf-net-performance}
\end{figure}

\paragraph{Three-objective front construction.}
Building on the stiffness--conductivity experiment, we further include volume fraction as an objective and use the network to densify a three-objective front for stiffness, thermal conductivity, and lightweight design. As shown in Fig.~\ref{fig:vf-net-performance}, the iterative search adds property--material-usage trade-off regions that were not covered by the initial random exploration. The final combined set achieves $HV=0.0523$, demonstrating that the proposed network-guided strategy remains effective in improving front coverage in the three-objective setting.

\subsection{Effectiveness of the Manufacturing Constraints}
\begin{figure}[t]
\centering
\includegraphics[width=\linewidth]{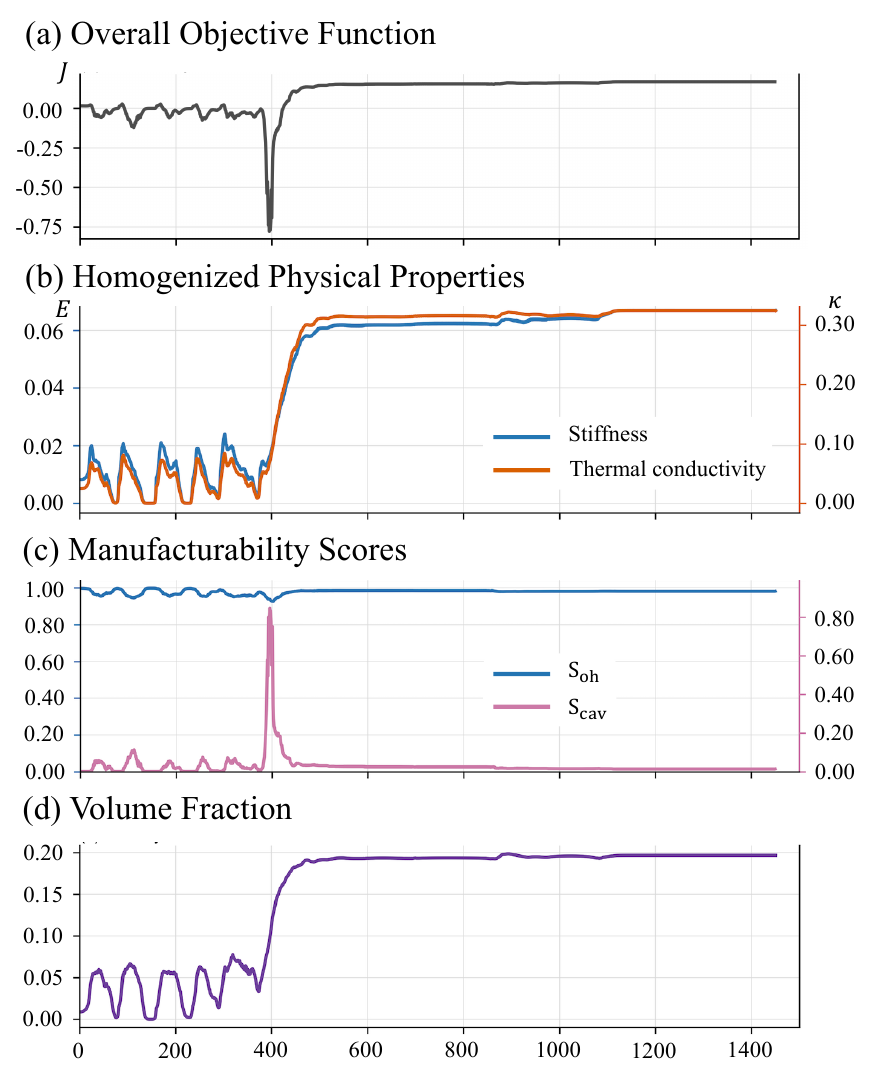}
\caption{\textbf{Iteration history of manufacturability-constrained topology optimization.} (a) Total objective $J$; (b) effective stiffness $E^H$ and thermal conductivity $\kappa^H$; (c) self-support score $S_{\rm oh}$ and cavity score $S_{\rm cav}$; and (d) physical volume fraction. The cavity score briefly fluctuates and then drops, the self-support score rises, and all measures subsequently stabilize.}
\label{fig:iter}
\end{figure}

We compare optimization with and without manufacturing constraints. The unconstrained formulation contains only the performance objective, whereas the constrained formulation adds the overhang and channel-aware cavity penalties. Condition sampling, volume fraction, initialization, and maximum iteration count are identical.

Figure~\ref{fig:iter} shows the iteration history of a representative constrained run. To visualize the manufacturability of intermediate candidates, the figure reports the binary-geometry scores $S_{\rm oh}$ and $S_{\rm cav}$ rather than the continuous penalties $P_{\rm oh}$ and $P_{\rm cav}$ used in the objective. The continuous terms provide differentiable guidance during density optimization; the binary scores directly describe overhang risk, enclosed cavities, and restricted powder-removal channels in the current geometry. As continuous load-bearing and heat-transfer paths emerge, effective stiffness and conductivity increase, and the physical volume fraction settles near its target. During the middle iterations, a change in local void connectivity briefly raises $S_{\rm cav}$ and reduces the total objective. The nonmanufacturable feature is subsequently removed, causing $S_{\rm cav}$ to drop rapidly, while $S_{\rm oh}$ increases and stabilizes. The total objective, physical properties, and both manufacturing scores then converge, showing that performance and manufacturability can be improved coherently within one optimization process.

\begin{figure}[t]
\centering
\includegraphics[width=\linewidth]{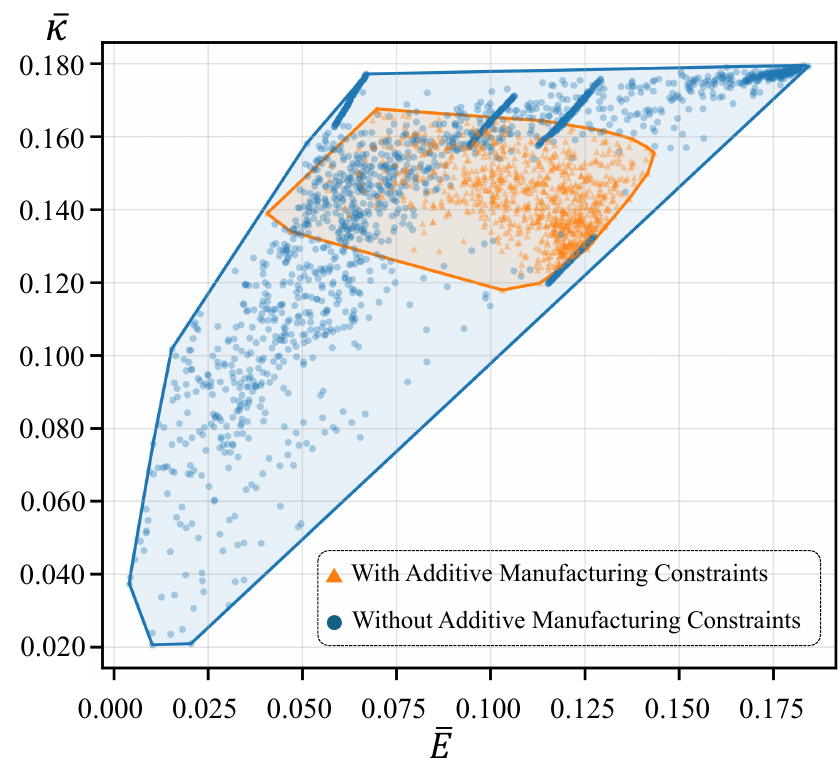}
\caption{\textbf{Effect of manufacturing constraints on the candidate performance space.} Orange and blue points denote candidates obtained with and without manufacturing constraints, respectively; each outer contour shows the corresponding objective-space coverage. The constraints contract the attainable performance space but confine candidates to a region that better satisfies practical AM requirements.}
\label{fig:fab-compare}
\vspace{-2mm}
\end{figure}

\begin{table}[t]
\centering
\caption{\textbf{Effect of manufacturing constraints on optimization and AM feasibility}}
\label{tab:am_constraint_compare}
\scalebox{0.94}{
\begin{tabular}{c|rrccr}
\hline
Method
& $\overline I\downarrow$
& $R_{\rm opt}\uparrow$
& $\overline S_{\rm oh}\uparrow$
& $\overline S_{\rm cav}\downarrow$
& $R_{\rm feas}\uparrow$\\
\hline
Unconstrained & \bfseries{184} & \bfseries{100.00\%} & 0.48 & 0.56 & 0.53\% \\
AM-constrained & 1516 & 78.30\% & \bfseries{0.96} & \bfseries{0.03} & \bfseries{99.11\%} \\
\hline
\end{tabular}
}
\vspace{-3mm}
\end{table}

\begin{figure}[t]
\centering
\includegraphics[width=\linewidth]{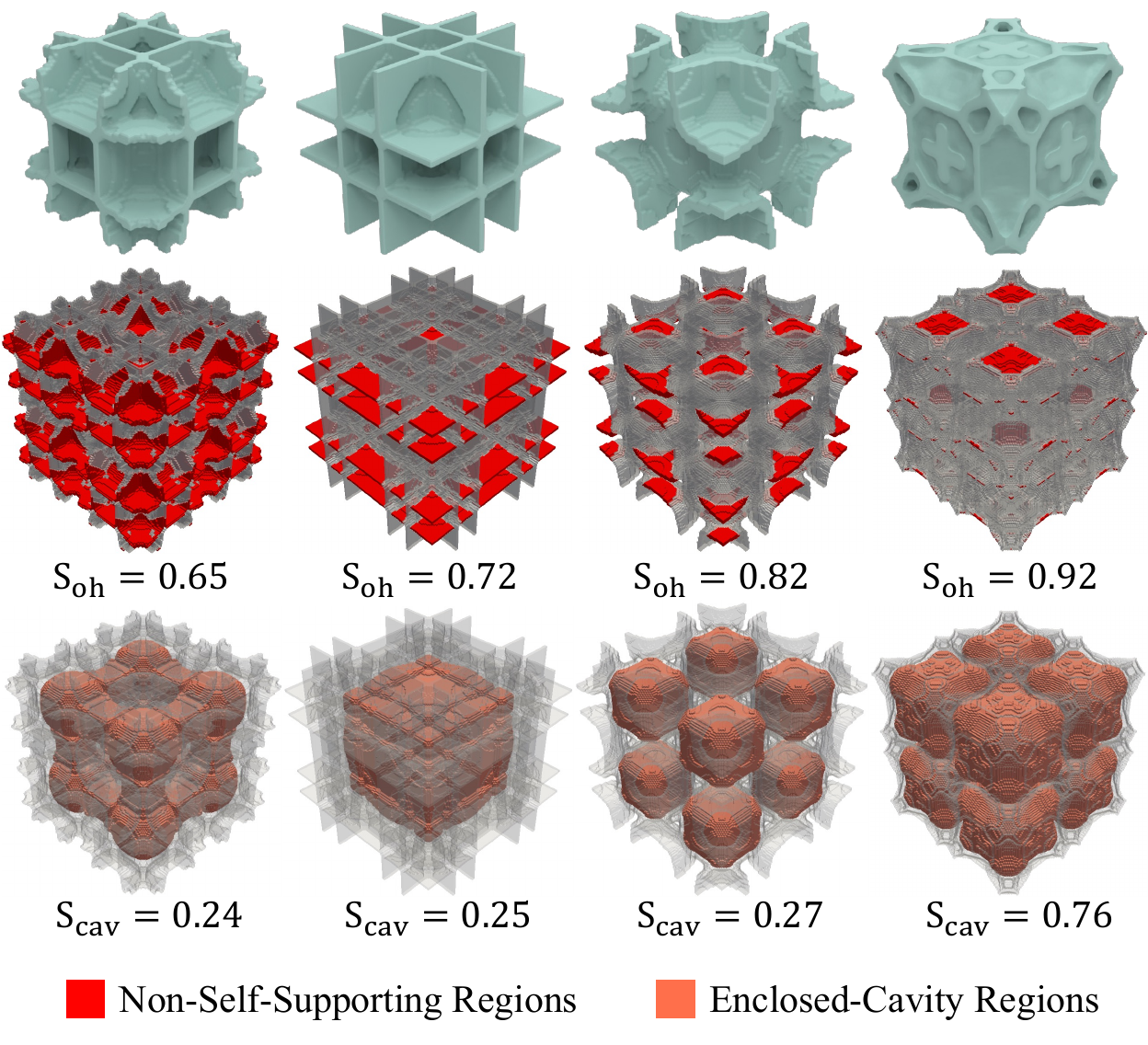}
\caption{\textbf{Representative results obtained without manufacturing constraints.} The first row shows optimized lattices. In the second row, red marks potentially unsupported solid and gray marks self-supported solid. In the third row, orange marks ineffective cavities or powder-removal regions and gray marks the remaining solid. Performance-only optimization readily produces overhangs, enclosed cavities, and restricted powder-removal paths.}
\label{fig:am_constraint_structures}
\end{figure}

\begin{figure}[t]
\centering
\includegraphics[width=.8\linewidth]{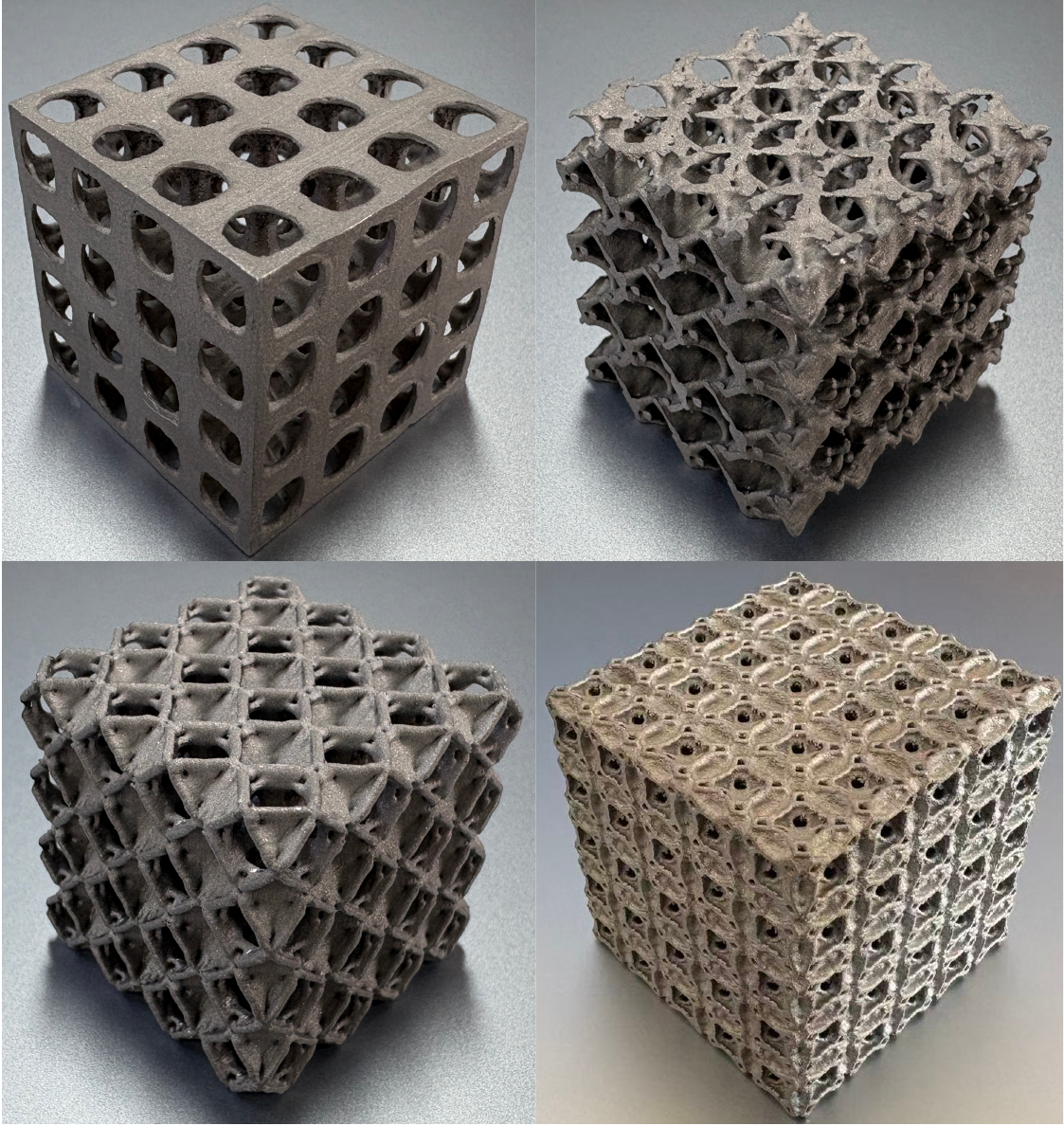}
\caption{\textbf{Metal prints of representative lattice structures optimized with manufacturing constraints.} The specimens exhibit well-connected solid networks and open pore channels.}
\label{fig:fab-results}
\end{figure}

Table~\ref{tab:am_constraint_compare} provides a quantitative comparison. The unconstrained formulation converges faster and succeeds more often because the optimizer can pursue the performance objective without manufacturing restrictions. Its structures, however, are only numerically valid and are rarely manufacturable. Adding the constraints increases the iteration count and slightly lowers the success rate, but substantially improves self-support, reduces cavity risk, and sharply increases the manufacturable fraction.

Figure~\ref{fig:fab-compare} compares the $\bar E$--$\bar\kappa$ distributions. The unconstrained formulation spans a larger performance region and contains extreme high-performance structures, but many violate manufacturing requirements. Constraints shrink the attainable region and remove these extreme topologies, while retaining a well-covered nondominated set within the manufacturable region. The constraints therefore confine the search to an engineering-feasible domain without eliminating the ability to construct a Pareto front. Figure~\ref{fig:am_constraint_structures} shows representative unconstrained designs, and Fig.~\ref{fig:fab-results} shows metal prints of selected manufacturable designs. These prints verify that several optimized geometries can be fabricated under the reported process settings; they do not by themselves validate the predicted effective properties.

\paragraph{Comparison with existing manufacturing treatments.}
We attempted to embed the support-free filter of~\cite{rosnitschek2023manufacturing} into our inverse-homogenization framework as a baseline, but it did not yield valid results in the present setting. That method performs a binary layerwise support test and removes unsupported material by hard deletion. Our optimization instead begins from a random continuous density that satisfies the volume constraint. Thresholding such a field creates many isolated floating solids, so the first filtering pass removes a large amount of material and collapses the structure into a sparse skeleton. Because this operation is discontinuous and nondifferentiable, material reintroduced by volume correction may be deleted again, causing the volume constraint to conflict with the support-free filter and preventing convergence to a valid lattice. A deletion-based binary filter is therefore difficult to apply directly to the continuous-density lattice problem considered here.

For enclosed cavities, the method of~\cite{wang2022topology} mainly creates powder-removal channels that connect existing closed cavities to the exterior and is therefore closer to post-optimization geometric compensation. It detects cavities and creates explicit channels on a binary geometry, so it cannot provide stable differentiable feedback to a continuous-density optimizer. Moreover, the boundary of a periodic unit cell is not necessarily a true exterior outlet, and connectivity within an isolated cell does not guarantee effective powder removal from a periodic array. Post-processed channels can also alter topology, volume fraction, and effective physical properties, making them difficult to coordinate with multiphysics objectives and Pareto-front construction. Although this approach can improve powder removal for a prescribed model, it does not directly address cavity and minimum-channel constraints during continuous optimization of periodic lattices.

\subsection{Parameter Sensitivity}

We vary $\lambda_{\rm oh}$ and $\lambda_{\rm cav}$ to study the strength of the manufacturing penalties. Twenty random initial density fields are generated in advance and shared by every parameter combination; all other settings remain fixed. Table~\ref{tab:lambda_sensitivity} summarizes the results.

For small penalty weights, the performance objective dominates: $\bar E$ and $\bar\kappa$ are high, but self-support is poor and cavity risk is substantial. Increasing $\lambda_{\rm oh}$ suppresses unsupported regions and raises $S_{\rm oh}$, whereas increasing $\lambda_{\rm cav}$ reduces enclosed cavities and lowers $S_{\rm cav}$. Excessive penalties, however, contract the performance region and reduce both $\bar E$ and $\bar\kappa$.

Considering all metrics, we use $\lambda_{\rm oh}=1$ and $\lambda_{\rm cav}=1$ as the default. This combination substantially improves manufacturability while preserving competitive physical performance.

\begin{table}[t]
\centering
\caption{\textbf{Performance and manufacturability for different penalty weights.}}
\label{tab:lambda_sensitivity}
\scalebox{0.94}{
\begin{tabular}{cc|rccccr}
\hline
$\lambda_{\rm oh}$
& $\lambda_{\rm cav}$
& $R_{\rm opt}\uparrow$
& $\bar E\uparrow$
& $\bar\kappa\uparrow$
& $\overline S_{\rm oh}\uparrow$
& $\overline S_{\rm cav}\downarrow$
& $R_{\rm feas}\uparrow$ \\
\hline
\multirow{3}{*}{0}
& 0 & \bfseries{100\%} & \bfseries{0.120} & \bfseries{0.135} & 0.69 & 0.35 & 0.00\% \\
& 0.5 & \bfseries{100\%} & 0.095 & 0.120 & 0.72 & 0.12 & 0.00\% \\
& 1 & 80\% & 0.092 & 0.110 & 0.77 & 0.05 & 0.00\% \\
\hline
\multirow{3}{*}{0.5}
& 0 & \bfseries{100\%} & 0.100 & 0.128 & 0.76 & 0.22 & 0.00\% \\
& 0.5 & 90\% & 0.088 & 0.113 & 0.88 & 0.10 & 11.11\% \\
& 1 & 85\% & 0.080 & 0.108 & 0.87 & 0.02 & 17.64\% \\
\hline
\multirow{3}{*}{1}
& 0 & 90\% & 0.091 & 0.118 & \bfseries{0.95} & 0.20 & 27.78\% \\
& 0.5 & 90\% & 0.085 & 0.110 & 0.90 & 0.05 & 55.56\% \\
& 1 & 75\% & 0.085 & 0.098 & 0.92 & \bfseries{0.01} & \bfseries{100.00\%} \\
\hline
\end{tabular}
}

\end{table}

\section{Conclusion}

We have presented a manufacturability-constrained lattice optimization framework for additive manufacturing and progressive Pareto-front construction. The core idea is to embed differentiable manufacturing constraints into inverse-homogenization topology optimization so that physical performance and manufacturability are considered within the same process. Pareto-front construction is driven by a closed loop between front evolution and model updating: latent representations of neighboring nondominated solutions are interpolated and decoded into initial fields for topology optimization, while iterative network updates coordinate front evolution with model learning.

Experiments demonstrate that the framework produces a well-distributed set of nondominated solutions under manufacturing constraints and reduces the risks associated with overhangs, enclosed cavities, and restricted powder removal. The iterative front-evolution strategy continually expands the nondominated solution space, providing a broader set of candidates for multiobjective lattice design. In addition, we constructed and organized a multiphysics lattice-structure optimization dataset for additive manufacturing, providing a foundation for subsequent studies of data-driven lattice design and manufacturability analysis.

Several limitations remain. First, the experiments are based primarily on a single base material, and the effects of material parameters for different metals on physical performance, manufacturing-constraint strength, and Pareto-front shape have not yet been investigated systematically. Second, the current manufacturing constraints focus on geometric manufacturability, including overhangs, enclosed cavities, and minimum powder-removal channels. Process-physics effects such as thermal distortion, residual stress, surface roughness, and build defects are not considered. Third, cavity and powder-channel detection relies on external-reachability analysis over a periodic supercell. This analysis requires explicit replication of the lattice unit cell and morphological propagation over a larger voxel domain, and its GPU memory consumption limits its applicability at higher resolutions. A hybrid CPU--GPU strategy that uses both system and GPU memory may help alleviate this limitation. Fourth, the current multiobjective solution strategy scalarizes multiple performance objectives through convex weighting. The effective stiffness, thermal conductivity, and shear modulus considered here are all obtained from homogenization problems governed by similar elliptic equations, and the experimental Pareto fronts exhibit no pronounced nonconvex regions; weighted scalarization therefore provides effective coverage in the present experiments. However, when the framework is extended to combinations of objectives governed by more disparate physical mechanisms, the Pareto front may become nonconvex, and weighted scalarization has theoretical limitations in recovering such regions~\cite{das1997closer}. Alternative decomposition strategies would then be required. Finally, the current density generation network is a regression model with limited capacity to represent large-scale Pareto-front distributions. Introducing generative models may further improve its modeling capacity.

\bibliographystyle{ACM-Reference-Format}
\bibliography{example}
\appendix
\section{Homogenization Theory}
\label{app:homogenization-theory}
\renewcommand{\theequation}{A\arabic{equation}}
\setcounter{equation}{0}

This appendix presents the theoretical basis and finite-element discretization used to homogenize a periodic lattice unit cell. Let the periodic representative volume element be $\Omega\subset\mathbb{R}^3$, with volume $|\Omega|$ and periodicity vectors $\boldsymbol\Upsilon=(\Upsilon_x,\Upsilon_y,\Upsilon_z)$. The cell is discretized into $N_{\mathrm{res}}^3$ eight-node hexahedral elements with densities $\rho_e\in[\rho_{\min},1]$. Here $\rho_e=1$ denotes solid, $\rho_e=\rho_{\min}$ approximates void, and $\rho_{\min}$ prevents a singular stiffness matrix.

Within the density-based framework, the solid isotropic material with penalization (SIMP) model relates element properties to density:
\begin{equation}
\mathbb{C}_e(\rho_e)=\rho_e^p\mathbb{C}_0,
\qquad
\kappa_e(\rho_e)=\rho_e^p\kappa_0,
\end{equation}
where $\mathbb{C}_0$ and $\kappa_0$ are the elasticity tensor and thermal conductivity of the base material, respectively, and $p>1$ is the penalization exponent. Solving the unit-cell problems under periodic boundary conditions maps the microscopic density field $\rho$ to the macroscopic effective elasticity tensor $\mathbb{C}^H$ and thermal-conductivity tensor $\boldsymbol\kappa^H$.

\subsection{Linear-Elastic Homogenization}

The effective elasticity tensor $\mathbb{C}^H$ relates macroscopic strain to volume-averaged stress and is represented as a $6\times6$ matrix in Voigt notation. Let $\boldsymbol\varepsilon^{0(I)}$ be the $I$th unit macroscopic strain mode, where $I=1,\ldots,6$ corresponds to the $11$, $22$, $33$, $23$, $13$, and $12$ components. For each mode, the periodic displacement fluctuation $\boldsymbol\chi^{(I)}$ satisfies
\begin{equation}
\nabla\cdot
\left[
\mathbb{C}(\rho):
\left(
\boldsymbol\varepsilon^{0(I)}+
\boldsymbol\varepsilon(\boldsymbol\chi^{(I)})
\right)
\right]=\boldsymbol 0,
\qquad \text{in }\Omega,
\end{equation}
where $\boldsymbol\varepsilon(\cdot)$ is the infinitesimal-strain operator. The fluctuation obeys periodic boundary conditions
\begin{equation}
\boldsymbol\chi^{(I)}(\boldsymbol x)=
\boldsymbol\chi^{(I)}(\boldsymbol x+\boldsymbol\Upsilon_\alpha),
\qquad
\boldsymbol x\in\partial\Omega,
\quad \alpha\in\{x,y,z\},
\end{equation}
and a zero-mean constraint that removes rigid translation:
\begin{equation}
\left\langle\boldsymbol\chi^{(I)}\right\rangle_\Omega=
\frac{1}{|\Omega|}\int_\Omega
\boldsymbol\chi^{(I)}\,\mathrm d\Omega
=\boldsymbol 0.
\end{equation}

The weak problem is to find $\boldsymbol\chi^{(I)}$ such that, for every periodic test function $\boldsymbol v$,
\begin{equation}
\int_\Omega
\boldsymbol\varepsilon(\boldsymbol v):
\mathbb{C}(\rho):
\boldsymbol\varepsilon(\boldsymbol\chi^{(I)})
\,\mathrm d\Omega
=-
\int_\Omega
\boldsymbol\varepsilon(\boldsymbol v):
\mathbb{C}(\rho):
\boldsymbol\varepsilon^{0(I)}
\,\mathrm d\Omega.
\end{equation}

Eight-node hexahedral elements are used for finite-element discretization. Let $\boldsymbol N$ be the shape-function matrix and $\boldsymbol B$ the strain--displacement matrix:
\begin{equation}
\boldsymbol u_e=\boldsymbol N\boldsymbol d_e,
\qquad
\boldsymbol\varepsilon(\boldsymbol u)_e=
\boldsymbol B\boldsymbol d_e.
\end{equation}
The element stiffness matrix and equivalent load vector are
\begin{equation}
\boldsymbol K_e^{\mathrm{el}}=
\int_{\Omega_e}
\boldsymbol B^{\mathrm T}
\mathbb{C}_e(\rho_e)
\boldsymbol B\,\mathrm d\Omega,
\qquad
\boldsymbol f_e^{\mathrm{el}(I)}=-
\int_{\Omega_e}
\boldsymbol B^{\mathrm T}
\mathbb{C}_e(\rho_e)
\boldsymbol\varepsilon^{0(I)}\,\mathrm d\Omega.
\end{equation}
Assembly gives the global systems
\begin{equation}
\boldsymbol K^{\mathrm{el}}(\rho)\boldsymbol d^{(I)}=
\boldsymbol f^{\mathrm{el}(I)},
\qquad I=1,\ldots,6.
\end{equation}

After the six unit-cell problems are solved, the effective elasticity tensor follows from energy equivalence:
\begin{equation}
C^H_{IJ}=
\frac{1}{|\Omega|}\int_\Omega
\left(
\boldsymbol\varepsilon^{0(I)}+
\boldsymbol\varepsilon(\boldsymbol\chi^{(I)})
\right):
\mathbb{C}(\rho):
\left(
\boldsymbol\varepsilon^{0(J)}+
\boldsymbol\varepsilon(\boldsymbol\chi^{(J)})
\right)
\,\mathrm d\Omega.
\end{equation}
Its discrete form is
\begin{equation}
C^H_{IJ}=
\frac{1}{|\Omega|}\sum_{e\in\Omega}
\left(
\boldsymbol d_e^{0(I)}+\boldsymbol d_e^{(I)}
\right)^{\mathrm T}
\boldsymbol K_e^{\mathrm{el}}
\left(
\boldsymbol d_e^{0(J)}+\boldsymbol d_e^{(J)}
\right),
\end{equation}
where $\boldsymbol d_e^{0(I)}$ contains the nodal affine-displacement degrees of freedom induced by the unit macroscopic strain, and $\boldsymbol d_e^{(I)}$ contains the periodic fluctuation degrees of freedom.

Let $\boldsymbol S^H=(\boldsymbol C^H)^{-1}$ be the effective compliance matrix. The directional Young's moduli are
\begin{equation}
E_x^H=\frac{1}{S^H_{11}},
\qquad
E_y^H=\frac{1}{S^H_{22}},
\qquad
E_z^H=\frac{1}{S^H_{33}}.
\end{equation}
We use their mean as the scalar stiffness measure:
\begin{equation}
E^H=\frac{1}{3}\left(E_x^H+E_y^H+E_z^H\right).
\end{equation}
The scalar shear modulus is extracted similarly:
\begin{equation}
G^H=\frac{1}{3}\left(
\frac{1}{S^H_{44}}+
\frac{1}{S^H_{55}}+
\frac{1}{S^H_{66}}
\right).
\end{equation}

\subsection{Steady-State Thermal Homogenization}

The effective conductivity tensor $\boldsymbol\kappa^H\in\mathbb{R}^{3\times3}$ relates a macroscopic temperature gradient to the volume-averaged heat flux. Let $\boldsymbol e_i$, $i\in\{x,y,z\}$, be a unit temperature-gradient direction. The corresponding periodic temperature fluctuation $\psi^{(i)}$ satisfies
\begin{equation}
\nabla\cdot\left[
\kappa(\rho)
\left(\boldsymbol e_i+\nabla\psi^{(i)}\right)
\right]=0,
\qquad \text{in }\Omega.
\end{equation}
The periodic boundary condition and zero-mean constraint are
\begin{equation}
\psi^{(i)}(\boldsymbol x)=
\psi^{(i)}(\boldsymbol x+\boldsymbol\Upsilon_\alpha),
\qquad
\boldsymbol x\in\partial\Omega,
\quad \alpha\in\{x,y,z\},
\end{equation}
\begin{equation}
\left\langle\psi^{(i)}\right\rangle_\Omega=
\frac{1}{|\Omega|}\int_\Omega\psi^{(i)}\,\mathrm d\Omega=0.
\end{equation}

For every periodic test function $q$, the weak form is
\begin{equation}
\int_\Omega
\nabla q\cdot\kappa(\rho)\nabla\psi^{(i)}\,\mathrm d\Omega
=-
\int_\Omega
\nabla q\cdot\kappa(\rho)\boldsymbol e_i\,\mathrm d\Omega.
\end{equation}

The thermal problem uses the same voxel mesh as the elasticity problem. Let $\boldsymbol B^{\mathrm{th}}$ be the gradient matrix of the temperature shape functions. The element conductivity matrix and load vector are
\begin{equation}
\boldsymbol K_e^{\mathrm{th}}=
\int_{\Omega_e}
\left(\boldsymbol B^{\mathrm{th}}\right)^{\mathrm T}
\kappa_e(\rho_e)
\boldsymbol B^{\mathrm{th}}\,\mathrm d\Omega,
\quad
\boldsymbol f_e^{\mathrm{th}(i)}=-
\int_{\Omega_e}
\left(\boldsymbol B^{\mathrm{th}}\right)^{\mathrm T}
\kappa_e(\rho_e)\boldsymbol e_i\,\mathrm d\Omega.
\end{equation}
The global system is
\begin{equation}
\boldsymbol K^{\mathrm{th}}(\rho)\boldsymbol\theta^{(i)}=
\boldsymbol f^{\mathrm{th}(i)},
\quad i\in\{x,y,z\},
\end{equation}
where $\boldsymbol\theta^{(i)}$ contains the discrete temperature-fluctuation degrees of freedom.

Energy equivalence gives the effective conductivity tensor:
\begin{equation}
\kappa^H_{ij}=
\frac{1}{|\Omega|}\int_\Omega
\kappa(\rho)
\left(\boldsymbol e_i+\nabla\psi^{(i)}\right)\cdot
\left(\boldsymbol e_j+\nabla\psi^{(j)}\right)
\,\mathrm d\Omega,
\qquad i,j\in\{x,y,z\}.
\end{equation}
Its discrete form is
\begin{equation}
\kappa^H_{ij}=
\frac{1}{|\Omega|}\sum_{e\in\Omega}
\left(\boldsymbol t_e^{0(i)}+\boldsymbol\theta_e^{(i)}\right)^{\mathrm T}
\boldsymbol K_e^{\mathrm{th}}
\left(\boldsymbol t_e^{0(j)}+\boldsymbol\theta_e^{(j)}\right),
\end{equation}
where $\boldsymbol t_e^{0(i)}$ contains the nodal affine-temperature values associated with the unit gradient and $\boldsymbol\theta_e^{(i)}$ contains the fluctuation values.

The scalar conductivity measure is the mean over the three principal directions:
\begin{equation}
\kappa^H=
\frac{1}{3}\operatorname{tr}(\boldsymbol\kappa^H)
=\frac{1}{3}\left(
\kappa^H_{xx}+\kappa^H_{yy}+\kappa^H_{zz}
\right).
\end{equation}

\section{Network Architecture and Hyperparameters}
\label{app:network_para}
\renewcommand{\theequation}{B\arabic{equation}}
\setcounter{equation}{0}
\renewcommand{\thetable}{B\arabic{table}}
\setcounter{table}{0}

During iterative Pareto-front construction, the target-performance-conditioned density generation network predicts an initial lattice density from a design condition and a target location in performance space. The network does not replace inverse-homogenization topology optimization; it provides a high-quality initialization that improves expansion efficiency in sparsely covered front regions.

\subsection{Condition Vector and Network Output}

The network receives a condition vector of dimension $d_c=7$:
\begin{equation}
\left[
w_E,
w_G,
w_\kappa,
v_f,
\bar E,
\bar G,
\bar\kappa
\right]^{\mathrm T}
\in\mathbb{R}^7,
\end{equation}
where $w_E$, $w_G$, and $w_\kappa$ are the weights of Young's modulus, shear modulus, and thermal conductivity in the scalar objective and satisfy
\begin{equation}
w_E+w_G+w_\kappa=1,
\qquad
w_E,w_G,w_\kappa\ge0.
\end{equation}
The target volume fraction is $v_f$, and $\bar E$, $\bar G$, and $\bar\kappa$ are normalized performance measures. In the two-objective experiment, $w_G=0$ but the input remains seven-dimensional to preserve a common interface.

The encoder produces a latent code of dimension $d_z=256$, and the network outputs a continuous density field at a resolution of $32^3$:
\begin{equation}
G_\theta(\mathbf t),
\qquad
\hat\rho\in[0,1]^{1\times32\times32\times32}.
\end{equation}
After volume-fraction correction, this output initializes inverse-homogenization topology optimization.

\subsection{Network Architecture}

Table~\ref{tab:app-network} lists the network modules.

\begin{table}[htbp]
    \centering
    \caption{Architecture of the target-performance-conditioned density generation network}
    \label{tab:app-network}
    \small
    \scalebox{0.83}{
    \begin{tabular}{cccc}
        \toprule
        Module & Input size & Output size & Main operations \\
        \midrule
        Condition encoder & $7$ & $256$ & Linear, LayerNorm, SiLU \\
        Fully connected lift & $256$ & $128\times4^3$ & Linear, SiLU, Reshape \\
        Decoding stage 1 & $128\times4^3$ & $64\times8^3$ & Upsample, ConvBlock3D \\
        Decoding stage 2 & $64\times8^3$ & $32\times16^3$ & Upsample, ConvBlock3D \\
        Decoding stage 3 & $32\times16^3$ & $16\times32^3$ & Upsample, ConvBlock3D \\
        Output layer & $16\times32^3$ & $1\times32^3$ & $1\times1\times1$ Conv, Sigmoid \\
        \bottomrule
    \end{tabular}
    }
\end{table}

\subsection{Training Hyperparameters}

Table~\ref{tab:app-network-para} lists the principal training settings.

\begin{table}[htbp]
    \centering
    \caption{Network-training hyperparameters}
    \label{tab:app-network-para}
    \small
    \begin{tabular}{l|c}
        \toprule
        Hyperparameter & Value \\
        \midrule
        Condition dimension & $7$ \\
        Output density resolution & $32\times32\times32$ \\
        Base spatial resolution & $4\times4\times4$ \\
        Latent dimension & $256$ \\
        Base channel count & $128$ \\
        Decoder channels & $128,64,32,16$ \\
        Optimizer & AdamW \\
        Initial learning rate & $2.0\times10^{-4}$ \\
        Weight decay & $1.0\times10^{-4}$ \\
        Training epochs & $200$ \\
        Batch size & $4$ \\
        Validation fraction & $0.1$ \\
        \bottomrule
    \end{tabular}
\end{table}

\subsection{Iterative-Evolution Hyperparameters}

During front construction, the network and nondominated set are updated alternately. In each iteration, the network generates initial densities in sparsely covered front regions. The newly optimized samples are merged into the training set, after which the network is fine-tuned. Table~\ref{tab:app-pareto-para} lists the corresponding settings.

\begin{table}[htbp]
    \centering
    \caption{Hyperparameters for iterative Pareto-front evolution}
    \label{tab:app-pareto-para}
    \small
    \begin{tabular}{l|c}
        \toprule
        Hyperparameter & Value \\
        \midrule
        Initial training epochs & $200$ \\
        Initial learning rate & $2.0\times10^{-4}$ \\
        Fine-tuning epochs per iteration & $50$ \\
        Fine-tuning learning rate & $1.0\times10^{-4}$ \\
        Fine-tuning batch size & $4$ \\
        Fine-tuning weight decay & $1.0\times10^{-4}$ \\
        Validation fraction & $0.1$ \\
        \bottomrule
    \end{tabular}
\end{table}

\section{Topology-Optimization Implementation}
\label{app:TO}
\renewcommand{\theequation}{C\arabic{equation}}
\setcounter{equation}{0}

We use the density-based SIMP method together with periodic density filtering, Heaviside projection, and an optimality-criteria (OC) update. The implementation follows the filtering--OC framework of~\cite{sigmund200199}, uses the projection formulation of~\cite{wang2011projection}, and performs material interpolation after density filtering, as commonly used for three-dimensional periodic inverse homogenization~\cite{zhang2023optimized}.

\paragraph{Periodic density filtering and projection.}
Let $\rho^d_e\in[0,1]$ be the design variable. To avoid discontinuities across periodic unit-cell boundaries, the filter neighborhood is defined using periodic distance. The filtered density at voxel $e$ is
\begin{equation}
\bar\rho_e=
\frac{\sum_{j\in\mathcal N_e}w_{ej}\rho^d_j}
{\sum_{j\in\mathcal N_e}w_{ej}},
\qquad
w_{ej}=\max\left(0,r_f-d_{\rm per}(e,j)\right),
\label{eq:filter}
\end{equation}
where $d_{\rm per}(e,j)$ is the shortest distance between voxel centers in the periodic domain, $\mathcal N_e=\{j\mid d_{\rm per}(e,j)<r_f\}$, and $r_f$ is the filter radius. We use $r_f=2$ voxels.

A smooth Heaviside projection then sharpens the solid--void boundary:
\begin{equation}
\hat\rho_e=
\frac{
\tanh(\beta_{\rm proj}\eta)+
\tanh\left[\beta_{\rm proj}(\bar\rho_e-\eta)\right]
}{
\tanh(\beta_{\rm proj}\eta)+
\tanh\left[\beta_{\rm proj}(1-\eta)\right]
},
\end{equation}
where $\eta=0.5$ is the projection threshold and $\beta_{\rm proj}=8$ controls the steepness. The physical density used for analysis and manufacturability evaluation is
\begin{equation}
\rho_e^{\rm phys}=
\rho_{\min}+(1-\rho_{\min})\hat\rho_e,
\label{eq:solid-void}
\end{equation}
where $\rho_{\min}=10^{-5}$ prevents singular finite-element matrices. Equation~\ref{eq:filter} suppresses checkerboards and controls the minimum feature size, while Eq.~\ref{eq:solid-void} drives the field toward a clear solid--void distribution.

\paragraph{Material interpolation and sensitivity propagation.}
Both elasticity and heat-conduction analyses use $\rho^{\rm phys}$. SIMP interpolation gives
\begin{equation}
\mathbf C_e(\rho_e^{\rm phys})=
(\rho_e^{\rm phys})^p\mathbf C_0,
\qquad
\kappa_e(\rho_e^{\rm phys})=
(\rho_e^{\rm phys})^p\kappa_0,
\end{equation}
with $p=3$. The finite-element systems are assembled from these properties, after which the periodic homogenization problems, performance objective, and manufacturing penalties are evaluated.

Let $\Phi=-J$ be the equivalent minimization objective. Automatic differentiation and the chain rule give
\begin{equation}
\frac{\partial\Phi}{\partial\rho^d_j}
=
\sum_e
\frac{\partial\Phi}{\partial\rho_e^{\rm phys}}
\frac{\partial\rho_e^{\rm phys}}{\partial\bar\rho_e}
\frac{w_{ej}}{\sum_{k\in\mathcal N_e}w_{ek}}.
\end{equation}
The projection derivative is
\begin{equation}
\frac{\partial\hat\rho_e}{\partial\bar\rho_e}=
\frac{
\beta_{\rm proj}
\left[1-\tanh^2\left(\beta_{\rm proj}(\bar\rho_e-\eta)\right)\right]
}{
\tanh(\beta_{\rm proj}\eta)+
\tanh\left[\beta_{\rm proj}(1-\eta)\right]
}.
\end{equation}

\paragraph{OC update and termination.}
The design density is updated with the OC method under the volume constraint. Let
$d_e=\partial\Phi/\partial\rho^d_e$ and
$v'_e=\partial V/\partial\rho^d_e$. The update factor is
\begin{equation}
B_e=\max\left(\epsilon,-\frac{d_e}{\lambda v'_e}\right),
\end{equation}
where $\lambda$ is the Lagrange multiplier for the volume constraint. It is determined by bisection so that the updated physical density satisfies
$N_e^{-1}\sum_e\rho_e^{\rm phys}\le v_f$. The design-variable update is
\begin{equation}
\rho_e^{d,\mathrm{new}}=
\operatorname{clip}_{[0,1]}
\left(
\operatorname{clip}_{[\rho^d_e-m,\rho^d_e+m]}
\left(\rho^d_e B_e^q\right)
\right),
\end{equation}
where the move limit is $m=0.05$ and the damping exponent is $q=0.5$.

Optimization terminates when the objective change remains below $5\times10^{-4}$ for three consecutive iterations or when the iteration count reaches $I_{\max}=2000$. The final output consists of the physical density field, effective performance measures, and manufacturability scores.

\end{document}